\documentclass[preprint2]{aastex}

\usepackage{natbib}
\usepackage{amsmath}
\usepackage{dsfont}

\shorttitle{Hi-GAL: star formation in the third Galactic quadrant}
\shortauthors{Elia et al.}

\begin{document}

\title{The first Hi-GAL observations of the outer Galaxy: a look to star formation in the third Galactic quadrant in the 
       longitude range $216.5\degr \leq \ell \leq 225.5\degr $}

\author{D. Elia\altaffilmark{1}, S. Molinari\altaffilmark{1}, Y. Fukui\altaffilmark{2}, E. Schisano\altaffilmark{3,1}, 
L. Olmi\altaffilmark{4,5}, M. Veneziani\altaffilmark{3}, T. Hayakawa\altaffilmark{2}, M. Pestalozzi\altaffilmark{1},
N. Schneider\altaffilmark{6}, M. Benedettini\altaffilmark{1}, A. M. Di Giorgio\altaffilmark{1}, D. Ikhenaode\altaffilmark{7},  
A. Mizuno\altaffilmark{8}, T. Onishi\altaffilmark{9}, S. Pezzuto\altaffilmark{1}, L. Piazzo\altaffilmark{6}, 
D. Polychroni\altaffilmark{10}, K. L. J. Rygl\altaffilmark{1}, H. Yamamoto\altaffilmark{2}, Y. Maruccia\altaffilmark{11}}
\altaffiltext{1}{Istituto di Astrofisica e Planetologia Spaziali - INAF, Via Fosso del Cavaliere 100, I-00133 Roma, Italy}
\altaffiltext{2}{Department of Physics, Nagoya University, Furo-cho, Chikusa-ku, Nagoya 464-8602, Japan}
\altaffiltext{3}{Infrared Processing and Analysis Center, California Institute of  Technology, Pasadena, CA, 91125, USA}
\altaffiltext{4}{Osservatorio Astrofisico di Arcetri - INAF, Largo E. Fermi 5, 50125 Firenze, Italy}
\altaffiltext{5}{University of Puerto Rico, Rio Piedras Campus, Physics Dept., Box 23343,  UPR station, San Juan, Puerto Rico, USA}
\altaffiltext{6}{IRFU/SAp CEA/DSM, Laboratoire AIM CNRS, Université Paris Diderot, F-91191 Gif-sur-Yvette, France}
\altaffiltext{7}{DIET – Dipertimento di Ingegneria dell'Informazione, Elettronica e Telecomunicazioni, Universit\`a di Roma ‘La Sapienza’, via Eudossina 18, 00184 Roma, Italy}
\altaffiltext{8}{Solar-Terrestrial Environment Laboratory, Nagoya University, Furo-cho, Chikusa-ku, Nagoya 464-8601, Japan}
\altaffiltext{9}{Department of Physical Science, Osaka Prefecture University, Sakai, Osaka 599-8531, Japan}
\altaffiltext{10}{University of Athens, Department of Astrophysics, Astronomy and Mechanics, Faculty of Physics, Panepistimiopolis, 15784 Zografos,Athens, Greece}
\altaffiltext{11}{Dipartimento di Matematica e Fisica - Universit\`{a} del Salento, CP 193, 73100, Lecce, Italy}

\email{davide.elia@iaps.inaf.it}

\begin{abstract}
We present the first Herschel PACS and SPIRE photometric observations in a portion of the outer Galaxy 
($216.5^{\circ} \lesssim \ell \lesssim 225.5^{\circ}$ and $-2^{\circ} \lesssim b \lesssim 0^{\circ}$) as a part of the Hi-GAL survey. 
The maps between 70 and 500~$\mu$m, the derived column density and temperature maps, and the compact source catalog are presented.
NANTEN CO(1-0) line observations are used to derive cloud kinematics and distances, so that we can estimate distance-dependent 
physical parameters of the compact sources (cores and clumps) having a reliable spectral energy distribution, that we separate 
in 255 proto-stellar and 688 starless. Both typologies are found in association with all the distance 
components observed in the field, up to $\sim~5.8$~kpc, testifying the presence of star formation beyond the Perseus arm 
at these longitudes. Selecting the starless gravitationally bound sources we identify 590 pre-stellar candidates. Several 
sources of both proto- and pre-stellar nature are found to exceed the minimum requirement for being compatible with massive star 
formation, based on the mass-radius relation. For the pre-stellar sources belonging to the Local arm ($d\lesssim1.5$~kpc) we study the 
mass function, whose high-mass end shows a power-law $N(\log M) \propto M^{-1.0 \pm 0.2}$. Finally, we use a luminosity vs mass 
diagram to infer the evolutionary status of the sources, finding that most of the proto-stellar are in the early accretion phase (with some 
cases compatible with a Class~I stage), while for pre-stellar sources, in general, accretion has not started yet. 
\end{abstract}


\section{Introduction}\label{intro}

Looking at star formation across the Milky Way, a large variety of different conditions is encountered, in many cases leading
to observed different modalities. One of the most relevant differences is found in the amount of star formation activity
between the inner and the outer Galaxy. The smaller content of star formation in the outer Galaxy is mainly related to the
lower density of atomic and molecular hydrogen compared with inner Galaxy \citep{wou90}.

Nevertheless the outer Galaxy offers the chance to study the interstellar clouds and possible ongoing processes of star formation
with a lower degree of confusion, also close to the Galactic plane. Furthermore, reconstructing the structure of the velocity field 
from line observations is easier in this case, compared to the inner Galaxy, thanks to the lack of distance ambiguities at any observed 
radial velocity.

In this paper we present the first photometric study of a portion of the outer Galaxy surveyed by the Herschel Space Observatory
\citep{pil10} as part of the Hi-GAL key-project \citep{mol10a}, selecting a range of longitudes  ($216.5^{\circ} \lesssim \ell
\lesssim 225.5^{\circ}$) in the third Galactic quadrant (hereafter TGQ), covered also by a CO(1-0) line map obtained with the NANTEN
radiotelescope. This combination of FIR/sub-mm dust continuum maps at unprecedented resolution and sensitivity and the information
about kinematics provided by the line observations allow us to investigate the first stages of the star formation in this portion
of the Milky Way, by estimating and discussing in statistical sense the physical properties of the compact sources (cores and clumps)
detected in the Herschel maps. With this analysis we aim at shedding more light on the star formation in the TGQ, to which a
relatively little amount of literature has been devoted so far.

The first extensive surveys of the the TGQ were carried out by \citet{may88,may93}, at an angular resolution of 0.5$^{\circ}$ and
$8\arcmin.8$, respectively. They confirmed the warping of the disk at these longitudes \citep{bur86} and the presence of weak
emission from molecular cloud located at Galactocentric distances larger than $R=12$~kpc. The latter survey was then drawn on by
\citet{may97} to derive the properties of the clouds located beyond 2~kpc from the Sun.
A relevant finding of this analysis is that no grand design spiral pattern is seen in the distribution of the clouds on the
Galactic disk. The warp of the Galaxy in the TGQ has been observed also by \citet{wou90} and \citet{dob05} through the distribution
of the IRAS sources and of the visual extinction maps derived from the Digitized Sky Survey I, respectively.

To assess the spiral structure of the Galaxy in the TGQ other techniques have been also adopted. \citet{car05} and
\citet{moi06} have studied the distribution of young star clusters, tracing the Local, the Perseus and the Norma-Cygnus
(Outer) arms. This latter is not detected at these longitudes by \citet{nak06} based on the \citet{dam01} CO survey
\citep[which is taken, for most of the TGQ, from that of][]{may93}. Instead, the \citet{moi06} results were confirmed
by \citet{vaz08}, who attribute to the Outer arm a grand design feature extending from $\ell=190^{\circ}$ to $255^{\circ}$
(reaching heliocentric distances $d=6$ and 12~kpc at the two extreme longitudes, respectively), expanding the sample of
clusters of \citet{moi06} and combining this information with kinematics extracted from new CO observations (still
unpublished at present).

Focusing on the vicinity of the region analyzed in this work, the most important star-forming region is CMa R1, dominated
by the presence of the CMa OB1 association \citep{rup66} located at $d=1150$ pc, i.e. in the Local arm, and with an age
of $3\times10^6$~yr \citep{cla74} and of the Sh2-296 (S296 in the simplified notation we adopt hereafter) H\textsc{ii}
region \citep{sha59}. The population of young stars in CMa R1 has been studied by \citet{gre09} based on X-ray data, and 
its typical age has been estimated as older than $10^7$~yr. However, only a very small fraction of the investigated 
proto-stars lie in the area surveyed in the present paper.

This area was fully covered, instead, by the survey of molecular clouds in the Monoceros and Canis Major regions carried
out with the NANTEN telescope in the $^{13}$CO(1-0) line \citep{kim04}, spanning an area of 560 square degrees. The two
most massive giant molecular clouds found are the aforementioned CMa OB1 and G220.8-1.7 \citep{mad86}, this latter located
at $d=1050$ pc. These authors also studied the relation between the gas emission morphology and the star formation 
signatures have been observed in this region, as several IRAS young stellar object candidates and H\textsc{ii} regions 
\citep{sha59,bli82}. The low level of extinction ($A_V<3$) found by \citet{dob05} 
in the entire region, however, seems to exclude the possibility
of encountering massive star formation \citep[MSF hereafter, cf, e.g.,][]{kru08}. The western part of our survey corresponds, 
instead, to the north-eastern corner of G216-2.5 (or Maddalena's cloud, $d=2200$~pc), mapped in both CO(1-0) and  $^{13}$CO(1-0) 
by \citet{lee94}. These authors claim the non-star-forming nature for this cloud, partially controverted by more recent Spitzer 
observations of \citet{meg09}, while recognize the star forming character of a nearby filamentary cloud associated with the S287 
H\textsc{ii} region, fully covered by our observations.

In this framework, the new Hi-GAL observations in the crucial range $70-500~\mu$m will permit to obtain a more 
exhaustive description of the star forming activities or capabilities in this region. In particular, both the early
stages of the proto-stellar core evolution and the properties of their pre-stellar progenitors can be studied, being 
in both cases the peak of the cold dust emission lying in this range of wavelengths \citep[e.g.,][]{mol10b,eli10}.

Thanks to NANTEN CO(1-0) data, we are also able to estimate kinematic distances for those sources, needed 
to derive masses and luminosities and then to depict a possible evolutionary scenario for the region. This 
allows us to carry out a complete study of cold dust and gas properties at unprecedented resolution and
sensitivity in a significant portion of the outer Galaxy.

This analysis is structured and presented as follows. In Section \ref{obs} the details about Herschel and NANTEN observations are
reported, together with data reduction strategy. In Section \ref{overall}, and also in Appendix \ref{allmaps}, the obtained maps
are presented, and used to provide an overall description of the column density and temperature distributions in the surveyed area.
Starting from Section \ref{cores} the discussion revolves about the compact sources that are extracted from the Hi-GAL maps and from
the ancillary CO(1-0) observations through cloud decomposition techniques. In Section \ref{seds} the main properties of the Hi-GAL compact
sources are derived from photometric data by means of a modified black body fit and then discussed in the framework of the differences
between pre- and proto-stellar sources. In Section \ref{evol}, in particular, these quantities are exploited to situate the Hi-GAL
sources into an evolutionary picture. Finally, in Section \ref{concl} a summary of the obtained results is reported.

\section{Observations and data reduction}\label{obs}

\subsection{Hi-GAL observations}
The \emph{Herschel Infrared Galactic plane survey} \citep[Hi-GAL,][]{mol10a} is a Herschel open-time key-project that
initially aimed at covering the inner Galaxy ($|\ell|<60^{\circ},|b|<1^{\circ}$), and was subsequently extended to the
entire Galactic disc.

In this paper we publish the first four ``tiles'' observed in the TGQ as part of this survey, spanning
the range of Galactic longitudes $216.5^{\circ} \lesssim \ell \lesssim 225.5^{\circ}$ and the range of latitudes
$-2^{\circ} \lesssim b \lesssim 0^{\circ}$, because in this portion of the Galactic plane the Hi-GAL observations
do not straddle the latitude $b=0^{\circ}$, but follow the warped midplane.

The $2.3^{\circ} \times 2.3^{\circ}$ fields presented in this paper were observed between May 5 and 9, 2011.
As in the case of all the Hi-GAL survey, the maps were obtained with the Herschel fast scan parallel mode, 
i.e., using simultaneously PACS at 70 and 160~$\mu$m \citep{gri10} and SPIRE at 250, 350, and 500~$\mu$m 
\citep{pog10} at the scan speed of 60$\arcsec /$s, and performing two scan maps of each field, one with 
nominal coverage and the other with orthogonal coverage, in order to increase data redundancy. Due to 
the 21$\arcmin$ separation of the PACS and SPIRE focal plane footprints on the sky, the areas covered
by these two instruments result slightly different, namely shifted each other of this angular quantity.
The diffraction limits at the five bands are 5.0, 11.4, 17.6, 23.9, and 35.1\arcsec, respectively. 
However, due to the PACS on-board data averaging and the use of the fast speed mode a degradation of 
the PACS PSF is produced along the scan direction, up to 12.2 and 15.7\arcsec at 70 and 160~$\mu$m, 
respectively (PACS manual, available from the ESA Herschel Science Centre). The Hi-GAL observation 
name convention consists in naming the tiles using the approximated value of the central longitude, 
then in the following we will designate these four tiles as $\ell217$ \footnote{In this case, 
the value $\ell=218^{\circ}$ would be closer to the center of the map, but we
use the conventional name already reported in the Herschel Data Archive.}, $\ell220$, $\ell222$,
$\ell224$. Hereafter, we will call $\ell217-224$ the entire region composed by these four tiles
and analyzed in this paper.

The data reduction strategy followed the Hi-GAL pipeline described in \citet{tra11}. Briefly, 
we used scripts written in the Herschel interactive processing environment \citep[HIPE,][]{ott10}
to reduce archival data to the Level 1. The obtained time-ordered data (TODs) of each bolometer
were then processed further by means of dedicated IDL routines to remove spurious instrumental
effects and glitches due to cosmic rays, and maps were obtained using the FORTRAN code ROMAGAL
based on Generalized Least Square \citep[GLS: e.g.,][]{teg97} approach, able to reduce the influence
of the $1/f$ noise in the TODs. Since the GLS technique is known to introduce artifacts in the maps,
namely crosses and stripes in correspondence of the brightest sources, a weighted post-processing
of the GLS maps \citep[WGLS,][]{pia12} has been applied to finally obtain images in which artifacts
are removed or heavily attenuated.

The pixel sizes of the maps are 3.2\arcsec, 4.5\arcsec, 6.0\arcsec , 8.0\arcsec, and 11.5\arcsec at
70, 160, 250, 350, 500~$\mu$m, respectively. The astrometry of the maps has been checked comparing
the positions of several compact sources in the 70~$\mu$m maps  with those of their possible WISE
\citep{wri10} counterparts at 22~$\mu$m, and the calculated rigid offset applied for correction.
Finally, the absolute calibration has been definitively performed by applying a linear transform of
the maps adopting the coefficients determined by comparing Herschel with IRAS and Planck, following
\citet{ber10}.

\begin{figure*}
\centering
\includegraphics[clip,angle=90,width=7.5cm]{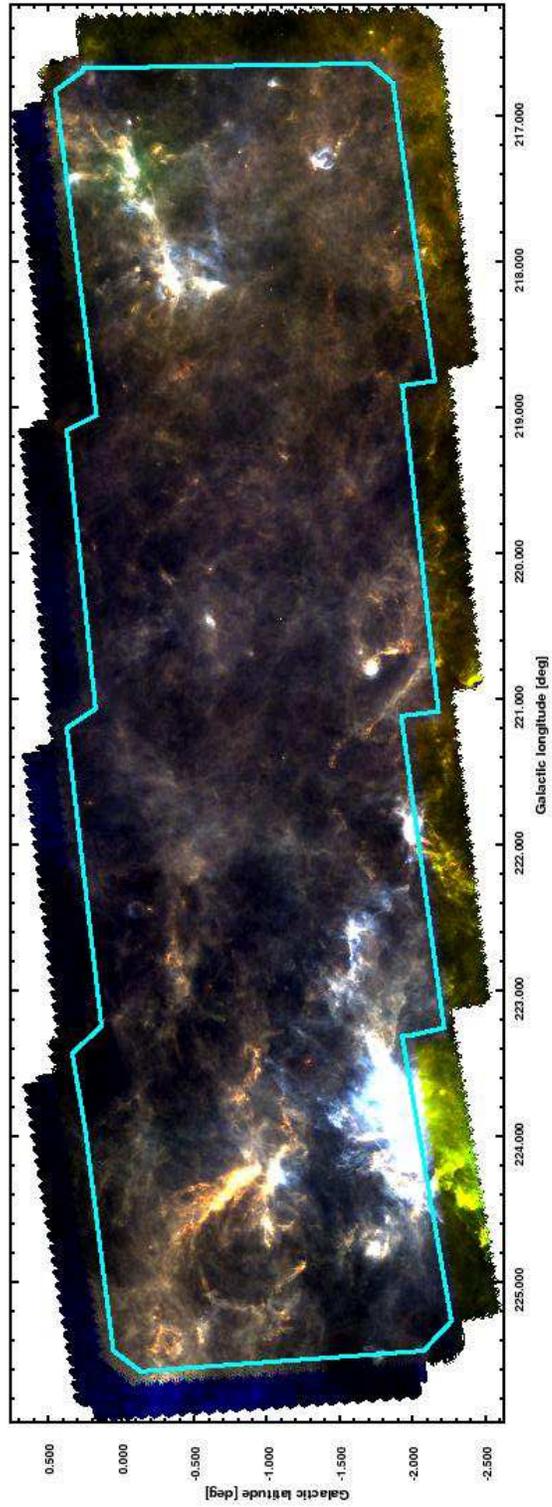}
\caption{RGB composite of $\ell217-224$ (blue: PACS 160~$\mu$m, green: SPIRE 250~$\mu$m, red: SPIRE 500~$\mu$m). The color
scaling is linear. The cyan contour delimits the common science area covered by both PACS and SPIRE.\label{rgbmosaic}}
\end{figure*}

All the single tile images obtained at the five Hi-GAL wavelengths are reported in APPENDIX \ref{allmaps}. In Figure~\ref{rgbmosaic},
instead, we show a composite RGB image of the region, assembled combining the mosaics of the four tiles at 160, 250 and 500~$\mu$m.
The 70~$\mu$m observations have not been used because the diffuse emission they contain appears dominated by the noise and only the
brightest regions are well recognizable. The mosaics have been produced in the same way as the single tiles, combining together the
TODs of the four tiles to obtain an overall map. The common science area, i.e. the area covered by PACS and SPIRE, highlighted with
a cyan contour in Figure~\ref{rgbmosaic}, amounts to $\sim 20.5$ deg$^2$.

\subsection{NANTEN CO(1-0) observations}

The NANTEN CO(J=1-0) Galactic Plane Survey data set \citep{miz04} was used in the present analysis. At the time of the observations, 
carried out in a period from 1996 to 2003, 4-m NANTEN millimeter-wave telescope of Nagoya University was installed at Las Camapans 
Observatory in Chile. The half-power beamwidth was $2\arcmin.6$. The data were gathered using the position switching mode.
The pointing accuracy was better than $20\arcsec$. The observations covered a region of $\sim 23^{\circ}$ (the coordinates of the 
bottom-left and top-right pixels are $[\ell,b]=[225.60^{\circ},-2.53^{\circ}]$ and $[\ell,b]=[216.53^{\circ},0^{\circ}]$, respectively), 
and the observed grid consists of $136 \times 38$ points located every $4\arcmin$. 

The spectrometer used was acousto-optical with 2048 channels providing a frequency resolution of 250~kHz. The spectral intensities 
were calibrated by employing the standard room-temperature chopper wheel technique \citep{kut81}. An absolute intensity calibration 
was achieved by taking the absolute peak antenna temperature, $T_{\mathrm R}^{*}$, of Orion KL (R.A. = $5^h32^m47.^s0$, 
Dec. = $-5^{\circ}24\arcmin 21\arcsec$ for the equinox of 1950.0) to be 65~K. The rms noise fluctuations were typically 0.2~ K 
with about 5 seconds integration for an on-position. A first order degree polynomial was subtracted from the line spectra to 
account for instrumental baseline effects. 
Finally, antenna temperatures were divided by the main beam efficiency factor $\eta_{\mathrm{mb}}=0.89$ \citep{oga90} to obtain
main beam temperatures.

The integrated intensity maps obtained from these line observations are presented in Section \ref{comaps_par}.

\section{The overall distribution of gas and dust}\label{overall}

\subsection{$^{12}$CO(1-0) line}\label{comaps_par}
A first look at the spectra in the reduced CO(1-0) data cube reveals that the structure of the velocity field is relatively clear due
to the small number of spectral components detected and to the infrequent presence of more than one along the same line of sight.

\begin{figure}
\epsscale{1.0}
\plotone{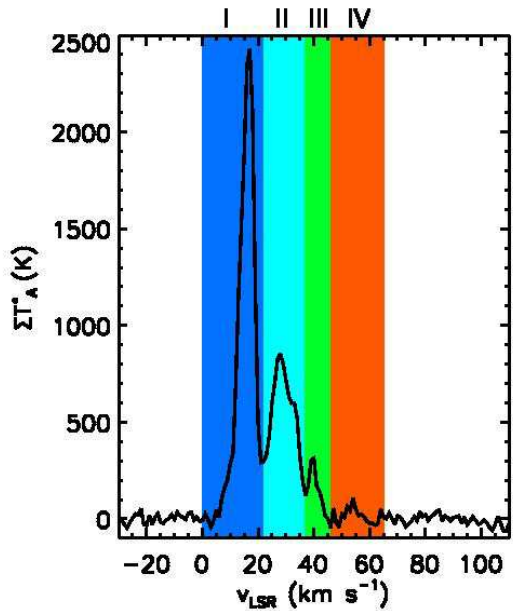}
\caption{Sum of all the $^{12}$CO(1-0) spectra. A rescaled $y$-axis is also reported on the right to give the  equivalent
interpretation as average spectrum. The four velocity ranges delimited by -0.5, 20.5, 36.5, 44.5, and 65.5~km~s$^{-1}$
are labelled with Roman numerals and highlighted with a blue, cyan, green and red background, respectively. \label{totspec}}
\end{figure}

A global view of the velocity field is provided in Figure~\ref{totspec} in which we show the sum of all the spectra composing the cube,
where at least three main components are found, peaking at the 17, 28, and 40~km~s$^{-1}$ velocity channels, respectively. A secondary
peak of the second component also appears around 33~km~s$^{-1}$. Finally, a further weak component, the most red-shifted, is present
around 54~km~s$^{-1}$. These indications can be exploited to determine opportune velocity ranges $[v_0,v_1]$ to build meaningful
channel maps, rather than blindly choosing equally spaced ranges of channels.

\begin{figure*}
  \includegraphics[width=13.5cm]{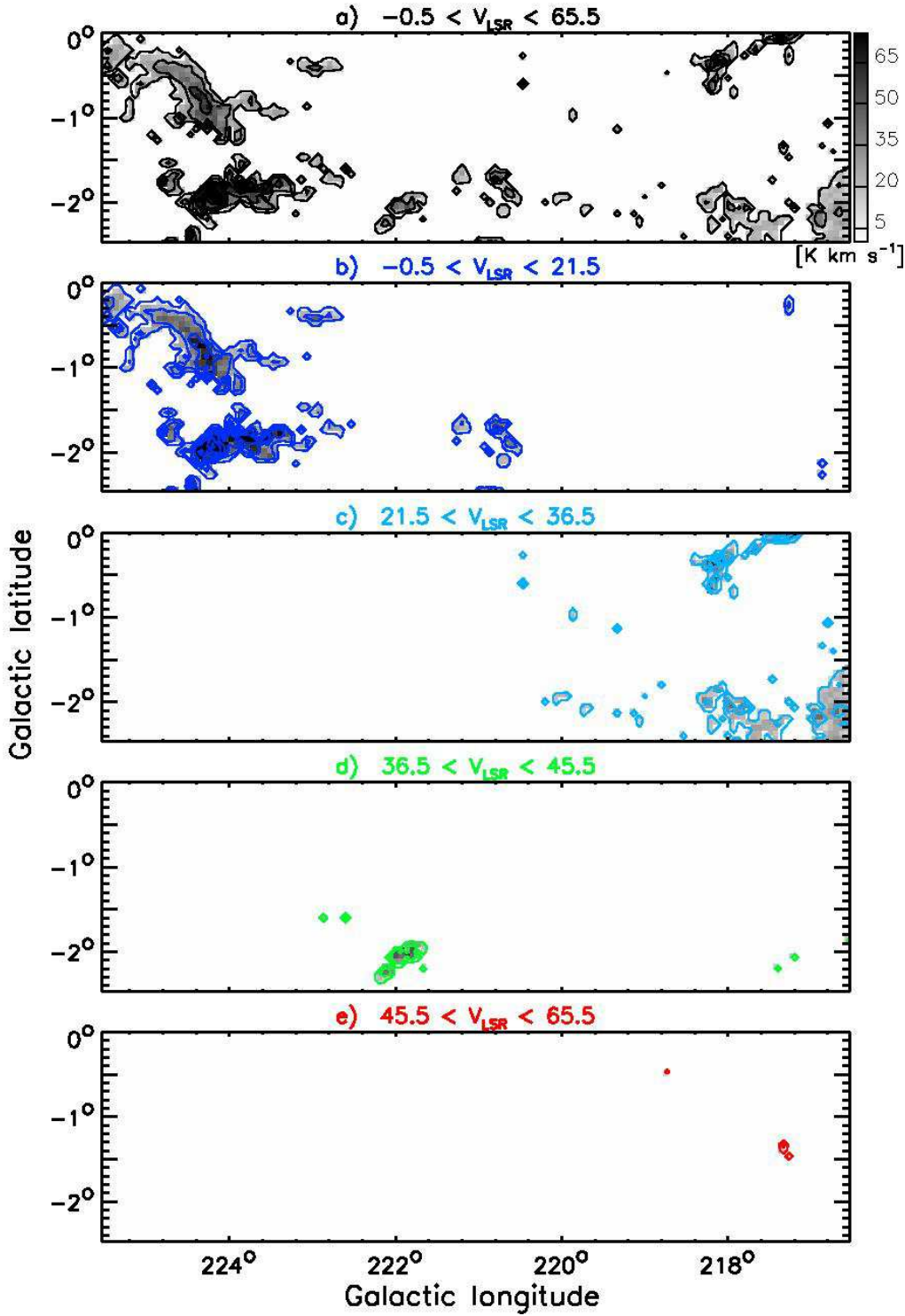}
  \caption{CO(1-0) integrated intensity maps of the $\ell217-224$ region, in units of K~km~s$^{-1}$. The correction for the
   $\eta_{\mathrm{mb}}$ factor is taken into account. In panel $a$, the map obtained integrating over in the entire considered
   velocity range is shown, while panels $b$-$e$ contain the intensity integrated over the ranges specified in the title of
   each panel. For all panels, contours start from 5~K~km~s$^{-1}$ and are in steps of 15~K~km~s$^{-1}$.
  \label{comaps}}
\end{figure*}

The maps have been obtained by integrating the line emission (main beam temperature) in those lines of sight where 
a feature with a signal-to-noise better than 3 has been found (the average baseline rms noise of the data cube 
is $\sigma_{\mathrm{rms}}=0.23$~K:
\begin{equation}
 I_{CO}=\int_{v_0}^{v_1}T^*_{\mathrm{mb}}\,dv
\end{equation}
In Figure~\ref{comaps} we show both the map of the $^{12}$CO(1-0) intensity integrated over the whole range from -0.5 to
60.5~km~K~s$^{-1}$ (panel $a$), and the set of the maps corresponding to the ranges highlighted in Figure~\ref{totspec}
(panels $b$-$e$).

\begin{deluxetable}{lcccccccc}
\tabletypesize{\scriptsize}
\tablecaption{Overall properties of the $^{12}$CO(1-0) emission in $\ell$217-224 \label{co_stats}}
\tablewidth{0pt}
\tablehead{
\colhead{Component} & \colhead{$v_{peak}$} & \colhead{$v_0$} & \colhead{$v_1$}& \colhead{Ref. longitude} &
\colhead{Distance} & \colhead{\% of total emission} & \colhead{$X$} &\colhead{Mass}\\
& \colhead{(km s$^{-1}$)} & \colhead{(km s$^{-1}$)} & \colhead{(km s$^{-1}$)}& \colhead{(deg)} &
\colhead{(kpc)} & \colhead{} & \colhead{$\times 10^{-20} (\mathrm{cm}^{-2}\, \mathrm{K}^{-1}\, \mathrm{km}^{-1}\, \mathrm{s})$} & \colhead{($10^5$ M$_{\odot}$)}\\}
\startdata
      I &        17 &    29.5 &    51.5 &       223 &       1.1 &   69.2 &    3.2 &    1.1  \\
     II &        28 &    51.5 &    66.5 &       218 &       2.2 &   26.2 &    3.3 &    1.9  \\
    III &        40 &    66.5 &    75.5 &       222 &       3.3 &    4.3 &    3.5 &    0.7  \\
     IV &        54 &    75.5 &    94.5 &       217 &       5.8 &    0.4 &    4.0 &    0.2  \\
\enddata
\end{deluxetable}

The first and the second velocity components (sorted from the bluest to the reddest) are evidently predominant in the total
budget of the emission. To quantify this, in Table~\ref{co_stats} the contribution of each component to the integrated intensity
map in panel~$a$ is quoted, together with the average distance calculated at a characteristic longitude for each component.
A gross estimate of the masses of these four components can be obtained, given the information on the global distance. First,
the local column density map is derived directly from the maps in Figure~\ref{comaps} through the empirical relation
 \begin{equation}\label{nh2co}
N(H_{2})_{i,j} = X\times {I_{CO}}_{i,j}\;,
\end{equation}
where the conversion coefficient value adopted in this case is calculated as a function of the Galactocentric radius $R$
through the relation of \citet{nak06}:
\begin{equation}\label{xco}
 X[\mathrm{cm}^{-2}\, \mathrm{K}^{-1}\, \mathrm{km}^{-1}\, \mathrm{s}]=1.4 \times 10^{20}\exp(R/\;11 \mathrm{kpc})\;.
\end{equation}

Then, for the each component the mass is obtained integrating the column density
over the area occupied in the sky: $M=\sum_{i,j} N(H_{2})_{i,j}\, d^2\, \Delta\vartheta^2\, \mu\, m_H$, where $\Delta\vartheta$ 
indicates the pixel scale of the CO(1-0) map (in radiants).

Nevertheless, the value of the $X$ conversion coefficient is matter of debate and strong variations of the obtained mass 
values are expected if another estimate of $X$ is adopted. For instance, \citet{nak06} make a twofold analysis using both the
approach described by the Equation \ref{nh2co} and the constant value $X=1.8 \times 10^{20}\mathrm{cm}^{-2}\, \mathrm{K}^{-1}\,
\mathrm{km}^{-1}\, \mathrm{s}$ of \citet{dam01}. In our case, this latter value would lead to underestimate of the masses 
Table~\ref{co_stats} of a factor from 1.8 to 2.2, depending on the distance.

\begin{figure*}
\includegraphics[width=17cm]{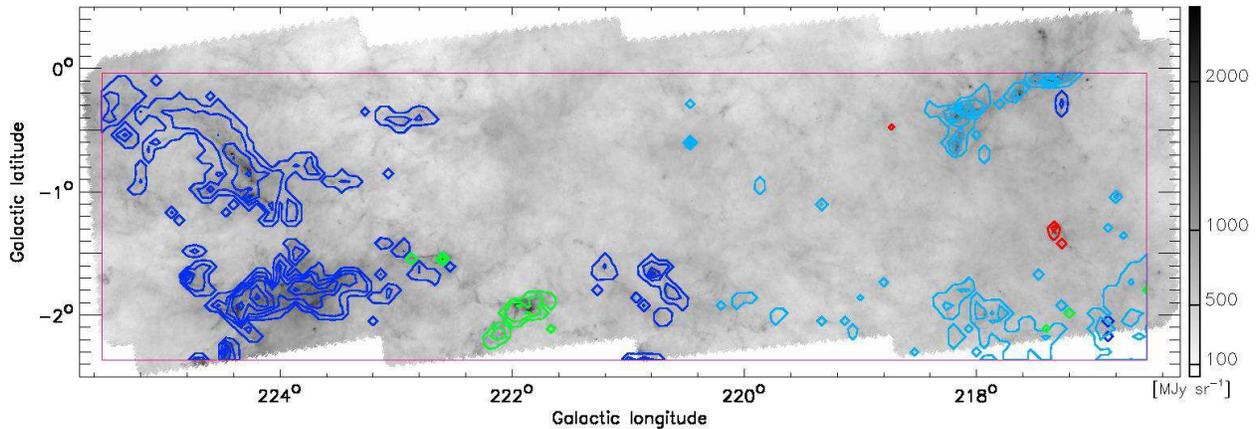}
\caption{Structure of the velocity field in $\ell217-224$. The contours of Figure~\ref{comaps}, panels $b$-$e$, are overplotted,
with the same color convention, on the the SPIRE 250~$\mu$m mosaic, while the magenta box represents the boundaries of the area
surveyed in the CO(1-0) line. The dynamics of the SPIRE image has been compressed between 20 and 2500~MJy/sr, and the scale is
logarithmic.
\label{vfield}}
\end{figure*}

The overall distribution of the column density derived from these maps is presented in Section~\ref{nh2_sect}, while the 
correspondence with Hi-GAL maps and with previous CO(1-0) surveys is discussed in Section~\ref{conh2}. Here we just highlight 
the low degree of complexity of the velocity field, as it emerges from these line observations. In Figure~\ref{vfield} the 
CO(1-0) contours of Figure~\ref{comaps}, $b-e$, are overlapped on the SPIRE 250~$\mu$m mosaic. The degree of overlap among components
appears relatively low: the eastern part of the region is dominated by the first (bluest) component, the western one is dominated by 
the second component, whereas the third and fourth component are basically associated to few small and bright features of the SPIRE map.

\subsection{Herschel temperature and column density maps} \label{nh2_sect}

Both the Herschel and NANTEN observations give the possibility of deriving maps of the column density in the region. Possible
differences between the resulting maps are expected, due to the assumptions made on the considered component of the ISM.
Therefore, any comparison requires care and must be accompanied with some caveats.

We obtained the maps of column density and temperature for $\ell217-224$ from the mosaics of the four Hi-GAL tiles, using 
the wavelengths from 160 to 500~$\mu$m, being the emission at 70~$\mu$m generally contaminated by warmer dust components as
circumstellar matter and H\textsc{ii} regions \citep[cf.][]{sch12}. This implies, of course, that these maps basically represent
the properties of the cold component of the dust (large grains). From the practical point of view, considering only the common
science area, all the maps have been rebinned onto the grid of the 500~$\mu$m one, and a pixel-to-pixel grey-body fit has been 
performed, according to the usual expression
\begin{equation}
\label{greybody}
F_{\nu}=N(H_2)\,\mu\,m_H\; \Delta\vartheta_{500}^2 \; k_0 \, \left(\frac{\nu}{\nu_0}\right)^{\beta}B_{\nu}(T)\;,
\end{equation}
where $F_{\nu}$ represents the spectral energy distribution (SED) of the pixel, $\Delta\vartheta_{500}$ is the 
map pixel scale at 500~$\mu$m in sr, $k_0=0.1$~cm$^2$~g$^{-1}$ at $\nu_0=1200$~GHz ($\lambda_0=250~\mu$m) already 
accounting for a gast-to-dust ratio of 100 \citep{hil83}, $m_H$ is the atomic hydrogen mass, and $\mu$ the mean molecular 
weight, assumed to be equal to 2.8 to take into account a relative helium abundance of 25\% in mass. 
We imposed $\beta=2$ in order to reduce the number of free parameters in the fit \citep[see also ][]{sad12}.

\begin{figure*}
\includegraphics[width=17cm]{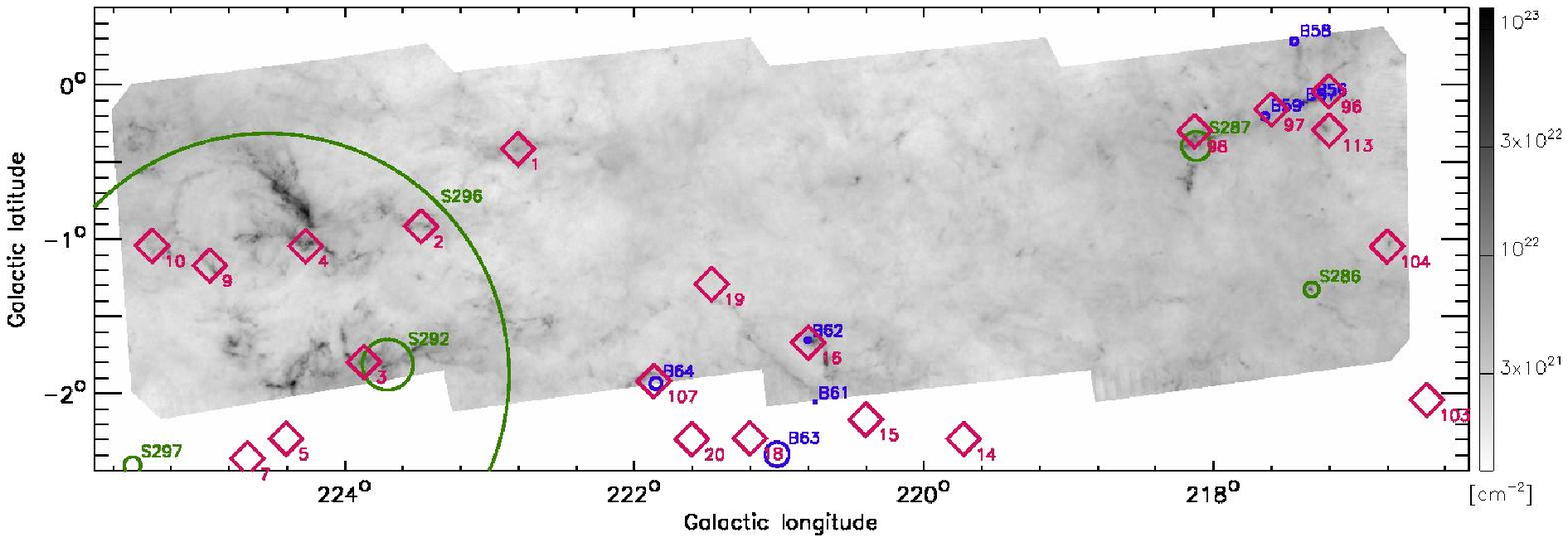}
\caption{Column density map of $\ell217-224$ in cm$^{-2}$, determined using the 160, 250, 350, 500~$\mu$m Herschel bands. 
The displacement of H\textsc{ii} regions is also shown: green circles are from \citet{sha59}, blue circles from \citet{bli82}, 
respectively. The object names, labeled at the top right of each region have been shortened to avoid
confusion in the figure (only those of BFS56 and BFS57 are partially overlapped in the North-West part of the map). Finally, the
molecular clouds of \citet{kim04} are overlapped as magenta diamonds, simply labeled with their running number at the bottom-right.
\label{hiiregions}}
\end{figure*}

\begin{figure*}
\includegraphics[width=17cm]{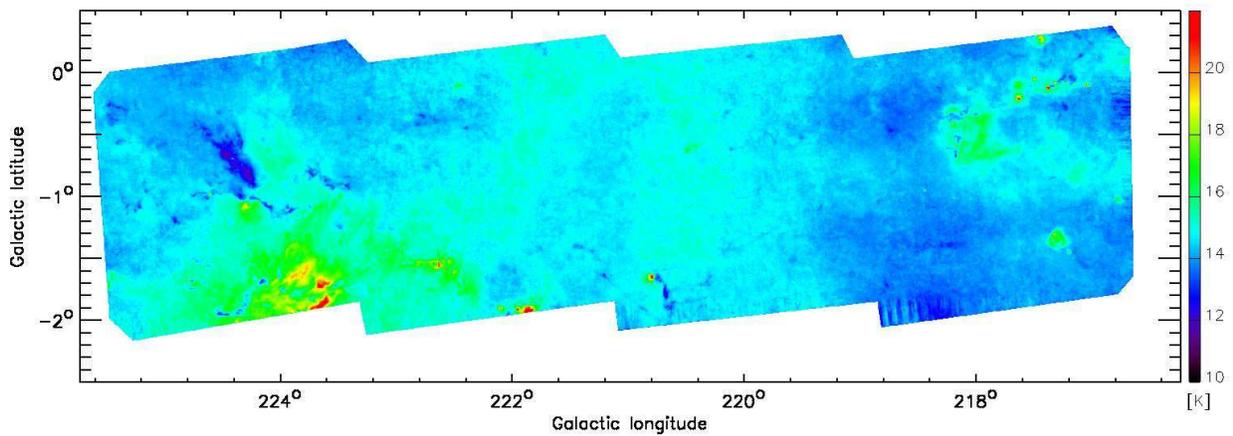}
\caption{Temperature map of $\ell217-224$, determined using the 160, 250, 350, 500~$\mu$m Herschel bands.
\label{htemp}}
\end{figure*}

The obtained column density and temperature maps are shown in Figures~\ref{hiiregions} and \ref{htemp}. Combining the information
contained in these maps and in that of Figure~\ref{vfield} with the global impression provided by Figure~\ref{rgbmosaic}, we are
now able to provide an overall description of the $\ell217-224$ region.

The eastmost tile, namely $\ell224$, is by far the richest in bright features and, apparently, in star formation activity. All 
these regions belong to the CMa~OB1 giant molecular cloud, located at $d=1150$~pc \citep[][and references therein]{kim04}, and 
are encircled by the H\textsc{ii} region S296 \citep{sha59}, as illustrated in Figure~\ref{hiiregions}.

A large cavity centered at about $[\ell,b]=[224.8^{\circ},-0.9^{\circ}]$ is well recognizable, with a filamentary ridge that turns
out to be particularly massive and cold ($11\lesssim T \lesssim 13$~K) in the western part. Another bright filament aligned with 
the East-West direction seems
to connect, in perspective, the center with the ridge of the cavity. This corresponds to the massive ($M=1.2\times 10^3 M_{\odot}$) 
cloud 4 of the paper of \citet{kim04} ([KKY2004]4 according to the SIMBAD nomenclature). The southern part of $\ell224$ is dominated 
by another extended, bright and warmer region corresponding to
S292 and [KKY2004]3; this latter is the most massive cloud ($M=1.9\times 10^3 M_{\odot}$) of those identified by \citet{kim04} in the
area considered in this paper. Cold filaments coexist with warm ($T<20$ K) diffuse emission or star forming sites, as S292. The young
cluster NGC2335, located north of this region ($[\ell,b]=[223.62^{\circ},-1.26^{\circ}]$) and with distance and age estimated in
1.79~pc and 79~Myr, respectively \citep{moi06} is not detected at any of the Herschel wavelengths.

In the $\ell222$ tile another small cloud is found, [KKY2004]1, still belonging to CMa OB1. In the southern part, the BFS64
H\textsc{ii} region \citep{bli82} is associated with the velocity component III. The emission in $\ell220$, again, is mostly
associated with the velocity component I. Two H\textsc{ii} regions, BFS61 and BFS62, represent the signature of recent 
high-mass star formation activity, albeit much less dramatic than in $\ell224$.

In $\ell217$ a filamentary cloud already mentioned in Section \ref{intro} is prominent in the northern part. It belongs to the velocity
component II and is associated with the S287 H\textsc{ii} region and other minor ones, and with four molecular clouds of \citet{kim04}.
The angular size of the whole filament is approximately 1.6$^{\circ}$, corresponding to an extent of 59~pc at a distance of
$d=2.1$~kpc. It is considered an active site of star formation by \citet{lee94}, who justify its filamentary shape with the action
of stellar winds of the O-B stars driving the H\textsc{ii} regions. A peak of dust temperature at $T=25.8$ K is found right by
the position of the BFS57 H\textsc{ii} region. The L1649 cloud \citep{lyn62}, indicated as [KKY2004]113, is well visible below
the filament, but physically unrelated with it, being instead the only prominent feature of the velocity component I in 
the $\ell217$ field. Finally, two dense and relatively compact regions are seen in the South. The first one corresponds 
to S286 and belongs to the velocity component IV ($d=5.3$~kpc, $R=10.0$~kpc), representing the farthest unconfused region 
detected in this portion of the Hi-GAL survey; in the dust temperature map it appears warmer ($15\lesssim T \lesssim 19$~K) 
than the portion of sky it is projected on. The second one (also IRAS06522-0350) is quoted as a satellite star-forming cloud 
\citep{meg09} of the G216-2.5 cloud \citep{lee94}.

As a global remark, we notice that, invoking the theoretical result of \citet{kru08} about a column density threshold for 
having massive star formation (MSF hereafter), namely $\Sigma=1$ g cm$^{-2}$ which assuming a mean molecular mass of 2.8 
corresponds to $N({\mathrm{H}_2}) = 2.1 \times 10^{23}$ cm$^{-2}$, there are no pixels exceeding this value in our column 
density map. However, this result cannot be considered resolutive against the presence of MSF in $\ell217-224$, for at 
least three main reasons: $i$) mainly for distant components, possible column density peaks might be diluted in the solid 
angle subtended by a pixel, because of the original instrumental angular resolution and/or the subsequent regridding of the 
maps for calculating the column density; $ii$) these column densities and temperatures are derived using the total emission 
measured along the whole line of sight; however, as it is described below in the paper, if the compact source 
emission is separated from the background, lower temperatures and higher masses are found for those cores, making it possible 
to find higher values of the column density; $iii$) as shown by \citet{kau10}, and also suggested by Hi-GAL observations 
\citep{eli10}, the \citet{kru08} critical value probably constitutes a too severe threshold for MSF; this point will be 
further discussed in Section \ref{higalprops}.

\subsection{Comparison of the two techniques} \label{conh2}

\begin{figure*}
\includegraphics[width=16cm]{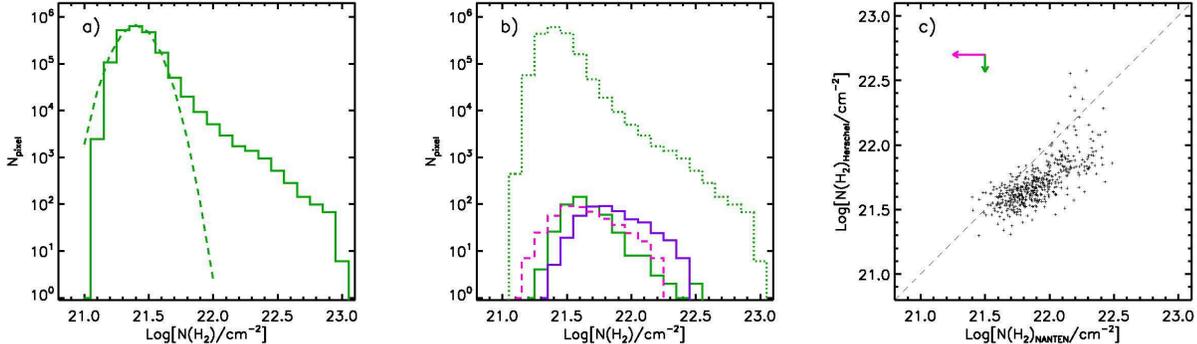}
\caption{Comparison between the Herschel- and CO-based column densities. Panel~$a$: Pixel distribution of the original 
Herschel-based map. The green dashed curve represents the log-normal fit to the Herschel histogram. 
Panel~$b$: The green dotted histogram is calculated as in panel~$a$, but considering only the pixels belonging to the area 
also covered by the CO(1-0) map. The green solid histogram is the pixel distribution of the Herschel-based column density map 
once reprojected on the CO map grid. The purple solid histogram is the distribution of the pixels of CO-based column density map 
belonging to the common coverage area. The magenta dashed histogram is the same distribution as the previous one, but calculated 
using the $X$ coefficient of \citep{dam01}.
Panel~$c$: Pixel-to-pixel plot of the values composing the dotted histograms of panel $a$. The dashed line represents 
the 1:1 relation. The green and magenta arrows represent the shift to be applied to the plotted points if different 
$k_0$ and $X$ coefficients are employed to compute the column density from the Herschel and the CO maps, respectively 
(see text).
The $x$ axes of all panels and the $y$ axis of the panel~$c$ are the same for comparison.
\label{histonh2}}
\end{figure*}

The availability of both the CO(1-0) and Herschel column density estimates allow us to compare the performance of an optically 
thick spectral line and of the dust as density tracers of the clouds, and to check the validity of the empirical relation~\ref{nh2co} in
this portion of the Galactic plane. We consider the Herschel-based map (Figure~\ref{hiiregions}) the most reliable of the two, being the 
cold dust a truly optically thin tracer of the ISM density. Instead, the empirical CO-based method may fail in single cases, mainly in 
the densest parts of the clouds, as recently further demonstrated by \citet{she11} by means of radiative transfer models. Also, it is 
sensitive to the excitation conditions of the gas. A systematic characterization of these factors affecting the CO-derived column 
densities will be reported in the forthcoming paper of \citet{car13}.

In Figure~\ref{histonh2}, panel~$a$, the histogram of this map is reported. It is characterized by a peak well described by 
a log-normal behavior (the log-normal fit is also shown), followed by a power-law-like extension at higher column densities. 
This situation is commonly found in regions forming low mass \citep{kai11} and high mass \citep{hil11,sch12} stars, being 
the log-normal shape attributed to turbulence, and the deviation at large densities due to strongly self-gravitating systems, a 
typical signature of star formation activity \citep[e.g., ][]{kle00,fed08}. A more quantitative analysis of
the power-law tail of this requires separating the different components along the line of sight. however, this is beyond
the scope of this present paper and will be presented in another study focusing on the probability density functions.

The CO-based column density has been derived for the four distance components through the Equation~\ref{nh2co}, adopting 
the corresponding $X$ factor values reported in Table~\ref{co_stats}; subsequently, these four maps have been co-added to obtain
the total column density map. To better compare the two maps, we considered only the portion corresponding to the common 
coverage area (Figure~\ref{histonh2}, panel~$b$, green dotted line), and then reprojected the Herschel-based map onto the grid 
of the CO(1-0) map, considering only the pixels with meaningful values in both maps (zeroes can be found, especially in 
the CO-based map). The histograms obtained in this way (Figure~\ref{histonh2}, panel~$b$, solid lines) become directly 
comparable, being derived from the same amounts of pixels.

It is possible to appreciate that the histogram of the CO-based column density appears quite different from those both the original 
Herschel-based map and its regridded version. In the first case, the largest values CO-based density values are more than half order 
of magnitude lower than those revealed in the Herschel-based map. This can be due to the combination of different concomitant
effects. First, the CO(1-0) is not ideal tracer of the high-density gas distribution. This mainly affects the histogram shape,
as a lack of the tail that, for instance, we observe in the Herschel-based one \citep[e.g.,][]{she11}. Second, the 
The significant pixel size difference between the SPIRE and NANTEN beams ($11.5\arcsec$ against $4\arcmin$) produces in the 
latter case a much higher degree of dilution of the local bright peaks in the final pixels, removing from the histogram the high-density 
tail that is seen instead in the 
former. This effect can be assessed by considering the histogram of the reprojected Herschel-based map: in this case the two 
histograms have a more similar peak position and ranges of variability. However, the regridding of the Herschel-based column density map 
produces a strong attenuation of both the low- and the high-density tail of its histogram, where now the CO-based one appears 
to be more populated. This can be seen also in the plot of the Herschel-based column density (reprojected) vs the CO-based one, built 
pixel-to-pixel and shown 
Figure~\ref{histonh2}, panel~$c$. In the majority of cases ($\sim 92\%$) the former is smaller than the latter and, the ratio of the 
sums of the involved pixels is $\sim 0.6$. This is just a global indication that cannot be simply taken as a sort of further 
correction of the adopted $X$ factors, since the relation between the two distributions does not appear linear. The $X$ factor is 
also known to depend strongly on local variations of the physical conditions, because it is expected to be lower in zones 
of star formation activity \citep{pin08}. Furthermore, the CO line intensity depends on the temperature of the gas. 
However, we can at least evaluate how the choice of the $X$ factor might affect the conclusions achievable by the comparison of 
the two column density maps. Adopting $X=1.8 \times 10^{20}\mathrm{cm}^{-2}\, \mathrm{K}^{-1}\, \mathrm{km}^{-1}\, \mathrm{s}$ 
\citep[][see Section~\ref{comaps_par}]{dam01}, the original CO map histogram is shifted towards lower densities, and its shape 
gets sligthly modified because of different coefficients initially adopted for the components I-IV (magenta dashed histogram 
in panel~$b$). In this case, the distribution peaks at a value smaller than that of the Herschel-based one. On the other hand, 
a different choice of the $k_0$ constant in the Equation~\label{greybody} would produce a rigid shift of the Herschel-based 
column density histogram. For example, the constant of \citet{pre93} ($k_0=0.005$~cm$^2$~g$^{-1}$ at 230~GHz) can be translated, 
for $\beta=2$ as in our case, to 0.14~cm$^2$~g$^{-1}$ at 1200~GHz, producing $40\%$ lower column densities. These possible 
corrections are represented in panel~$c$ as arrows, showing that the choice of a smaller $X$ coefficient would generally 
produce a better agreement between the two distributions, while a larger dust opacity would obtain the opposite effect.

\section{Cores and clumps}\label{cores}
In this section we start to focus on the distribution and the properties of the compact sources detected in the Hi-GAL
observations of $\ell217-224$. The distance estimate for these sources is performed exploiting the kinematic distances of
CO clumps.

\subsection{Photometry of Herschel compact sources}
The compact source detection and photometry have been performed at each of the five available bands following the procedure
described in previous Hi-GAL articles \citep{mol10b,eli10}. We used the Curvature Threshold Extractor package
\citep[CuTEx,][]{mol11}, which detects the sources as local maxima in the second-derivative images and then fits an elliptical
Gaussian to the source brightness profile to estimate the integrated flux of the source. The parameters of the best fit (the
peak position and strength, the minimum and maximum FWHM, and the integral of the Gaussian) are used to estimate the source
position, the peak flux, the angular size and the total flux. The flux uncertainties have been derived from the best fit as 
well. Using this approach, five independent lists of photometry have been obtained at the five Hi-GAL wavebands. 

\begin{figure}[ht]
\epsscale{1.1}
\plotone{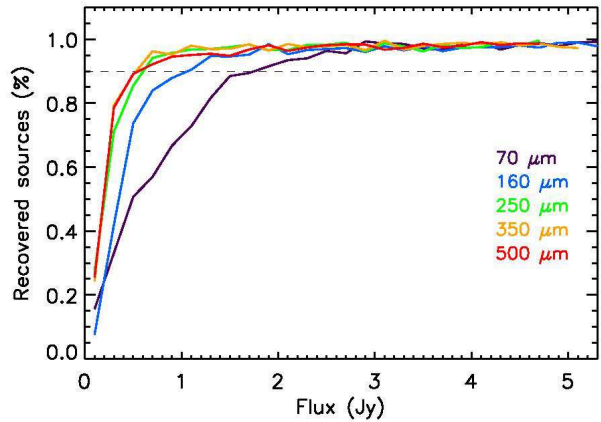}
\caption{Completeness limit calculation for Hi-GAL photometry. The curves represent the percentage of recovered artificial 
sources as a function of their fluxes in the five Herschel bands. The dashed horizontal line represents the 90\% limit.
\label{completeness}}
\end{figure}

The completeness limits have been estimated evaluating the CuTEx ability of recovering synthetic compact sources randomly
dispersed on the original maps. Both the size and the flux of the generated sources mirror the distributions contained in
the original CuTEx lists. The completeness limit at each waveband is then estimated as the flux value at which the histogram
of the fraction of the recovered sources exceeds the threshold of 90\% of the total number of injected synthetic sources 
having fluxes in that bin. The procedure is illustrated in Figure~\ref{completeness}, and the obtained values are 1.75, 
1.06, 0.60, 0.53, and 0.56 Jy at 70, 160, 250, 350, 500~$\mu$m, respectively. 

Subsequently the sources extracted in each waveband have been associated by means of a simple
spatial criterium \citep{eli10} in order to obtain a band-merged catalog. The total number of entries in this catalog is
3476, composed by sources having at least a detection in one band. In Section~\ref{seds} the further selection of a
sub-sample of reliable SEDs eligible for the fit through a model is reported.

\subsection{Cloud decomposition}\label{cdecomp}
Similarly to the source extraction performed on the Herschel maps, a search for spatially and kinematically coherent
overdense structures has been carried out in the CO(1-0) $l-b-v$ cube, using cloud decomposition techniques. Doing this
we do not aim at finding exactly the counterparts of the Herschel sources and at establishing a some biunivocal correspondence
between the two populations. Rather, due to the coarser spatial resolution of the CO(1-0) observations and to the different source
extraction techniques adopted, the CO(1-0) coherent sources (hereafter ``CO clumps'', although in some cases they could not
match exactly the definition of clump having typical cloud sizes) are expected to be generally more extended than the Herschel
ones, and thus to be found in association with more than one of them in general.

For the cloud decomposition we used the CLOUDPROPS package\footnote{The CLOUDPROPS package is available at
\url{https://people.ok.ubc.ca/erosolo/cprops/index.html}.} \citep{ros06,ros11}, which is also able to derive a set of
the basic morphological and physical properties of found the clumps. We tested both the native clump finding
algorithm present in the package, and the widely used cloud decomposition algorithm CLUMPFIND \citep{wil94},
that is also implemented in the package as alternative option. Although CLUMPFIND is known to be strongly depending on
the initial parameter setup, as discussed in detail, e.g., by \citet{bru03}, \citet{ros05}, \citet{ros06}, and
\citet{eli07}, in this case it turned out to be more able than the native CLOUDPROPS search engine to detect structures
with the same efficiency across the whole map, probably due to different rms noise conditions throughout the data cube
\footnote{In more detail, the native engine of CLOUDPROPS has not been able to detect evident structures in some
portions of the map even at very low detection thresholds, that included a number of false detections in other locations.
Instead, CLUMPFIND managed to recover all the main features of the data cube using a global and reasonable value of the
global threshold \citep[$3 \sigma$, see][]{wil94}}.

The list of the detected clumps is reported in Table~\ref{COclumps} in Appendix~\ref{clumplist}, complemented with the
following quantities provided by CLOUDPROPS and calculated as described in \citet{ros06}, with some further tuning of
the default settings:
\begin{itemize}
 \item \emph{Velocity dispersion}. It is derived as the second moment of the intensity distribution along the velocity
 dimension.
 \item \emph{Distance}. To derive the heliocentric distances, the Galactic rotation curve of \citet{bru11} ($R_0=8.3$~kpc,
 $\Theta_0=239$ km s$^{-1}$) has been used instead of the default one of \citet{rei09} implemented in CLOUDPROPS.
 \item \emph{Radius}. The deconvolved radius $r$ is estimated as the geometric mean of the deconvolved major and minor axes
 of the clump, given in units of pc once the clump distance is known.
 \item \emph{Mass}. The mass of the clump $M$ is obtained as in Section~\ref{comaps_par}, i.e. integrating the column density
 over the solid angle of the clump. The column density is derived by CLOUDPROPS as in Equation~\ref{nh2co} with a $X$
 factor constant for all clumps. We then rescaled the mass considering a $X$ factor varying with the Galactocentric
 distance as in Equation~\ref{xco}.
 \item \emph{Virial mass}. The virial mass is calculated as $M_{\mathrm{vir}}[M_{\odot}] = 1040 \sigma^2_v[\mathrm{km~s^{-1}}]~r[\mathrm{pc}] $,
 under the assumption that each clump is spherical and virialized with a density profile of the form $\rho \propto r^{-1}$
 \citep{sol87}. Density power laws with exponent 0 and -2 would require, instead, a constant of 1150 and 690 M$_{\odot}$
 (km~s$^{-1}$)$^{-1}$ pc$^{-1}$, respectively.
\item \emph{Luminosity}. The clump luminosity in the CO(1-0) line is provided in units of K~km~s$^{-1}$~pc$^2$.
 The conversion factor to solar luminosities for this transition is $4.9 \times 10^{-5} L_{\odot}
 [\mathrm{K~km~s^{-1}~pc^2}]^{-1}$ \citep{sol92}.
\end{itemize}

After ruling oututput sources having meaningless physical parameters (as not-a-number masses, luminosities, etc.) the final list
is composed by 321 clumps.

\section{Herschel sources: physical properties}\label{seds}

\subsection{SED building}\label{sedbuild}
In this section we derive the basic physical properties of the Hi-GAL compact sources from their spectral energy
distributions (SEDs). However, first of all a selection has to be performed in order to obtain a reliable sample
of SEDs eligible for the modified black body fit.

Where present, the flux at 70~$\mu$m is not considered in this fit, being generally affected by the direct contribution
of a proto-stellar component; for this reason, its presence can be used as a signature to distinguish a proto-stellar
from a starless core/clump \citep[e.g.][]{bon10,gia12}. We searched also for possible WISE 22~$\mu$m counterparts
(detected at $S/N>5\sigma$) within a searching radius of 12.6\arcsec (corresponding to the WISE resolution 
at that wavelength and to the WISE and Herschel positional uncertainties combined in quadrature) around 
the Hi-GAL source coordinates. The low signal-to-noise ratio in our PACS 70~$\mu$m maps, responsible of the large flux 
completeness limit at this band, can result in an association between a WISE source and a Herschel one lacking emission
at 70 $\mu$m. In these cases, the association is considered genuine and the source is classified as proto-stellar if 
at least the flux at 160~$\mu$m is available. In this way, 236 out of 943 Hi-GAL sources with a selected SED (see 
below) are found to be associated with a WISE counterpart.

To estimate which fraction of the WISE-Herschel associations could be spurious, we generated a random distribution
of coordinate pairs having the same number and spatial coverage of the WISE sources in our field, and run the same 
procedure of matching with the Hi-GAL sources. Only 18 of the aforementioned 943 Hi-GAL sources are associated with
a source generated in such a way, i.e. less than 8\% if compared with the number of the WISE counterparts found. This 
percentage represents the occurrence of spurious matches that should be expected to affect our WISE-Herschel association.

At the longer wavelengths, we deal with the frequent case of the increase of the beam-deconvolved diameter, as estimated by
CuTEx ($\theta_{\lambda} = \sqrt{FWHM_{maj,\lambda}\times FWHM_{min,\lambda}-HPBM_{\lambda}^2}$), at increasing wavelengths,
already discussed by \citet{mot10} and \citet{gia12}. Due to this increase of the area over which the emission is integrated,
many SEDs flatten or even rise longwards $\lambda \ge 250 \mu$m.

Here we adopt the strategy of \citet{mot10}, consisting in assuming a reference wavelength to estimate the source angular 
size, and scaling the fluxes at a larger wavelength $\lambda$ by the ratio between the deconvolved sizes at the reference
wavelength and at $\lambda$, respectively. Due to the further selection to identify SEDs eligible for the fit, described at 
the end of this section, the shortest wavelength always available is 250~$\mu$m, that we adopt as reference wavelength
\begin{equation}
 \overline{F}_{\lambda}=F_{\lambda}\times\frac{\theta_{250}}{\theta_{\lambda}}, \qquad \lambda \geq 350 \mu\mathrm{m}\;.
\end{equation}
This is based on the assumptions that $i$) the source is optically thin at $\lambda \geq 250\; \mu$m; $ii$) the
temperature gradient is weak \citep{mot01}; $iii$) the radial density profile of the source is described by
$\rho(r) \propto r^2$, then $M(r) \propto r$. 

It is noteworthy that scaling fluxes in this way corresponds to consider fluxes emitted by the same volume of dust,
that is more compatible with fitting a modified black body to a SED. On the other hand, hereafter it must be kept in
mind that the physical properties derived from the fit of the scaled SEDs describe the average conditions of the source inside
a volume delimited by the size detected at the reference wavelength. Finally, considering a smaller volume, we reduce the
influence of possible multiplicities (i.e. shorter wavelength counterparts that appear confused in an unresolved source at
350-500~$\mu$m) on the fluxes composing the final SED.

A further selection has been performed on the SEDs built as described above, accepting only those:
$i$) being composed by consecutive fluxes (at least three); $ii$) showing no dips (e.g. concave shape); $iii$)
not peaking at 500~$\mu$m; $iv$) having a distance estimate, as described in Section \ref{hdistance}.

After applying these constraints, we finally obtained a sample of 943 SEDs eligible for the fit, 255 of which are
proto-stellar (184 of which with detected at both 22~$\mu$m and 70~$\mu$m, 52 only at 22~$\mu$m, and 
19 only at 70~$\mu$m), and 688 starless.

\begin{figure*}
\includegraphics[width=17cm]{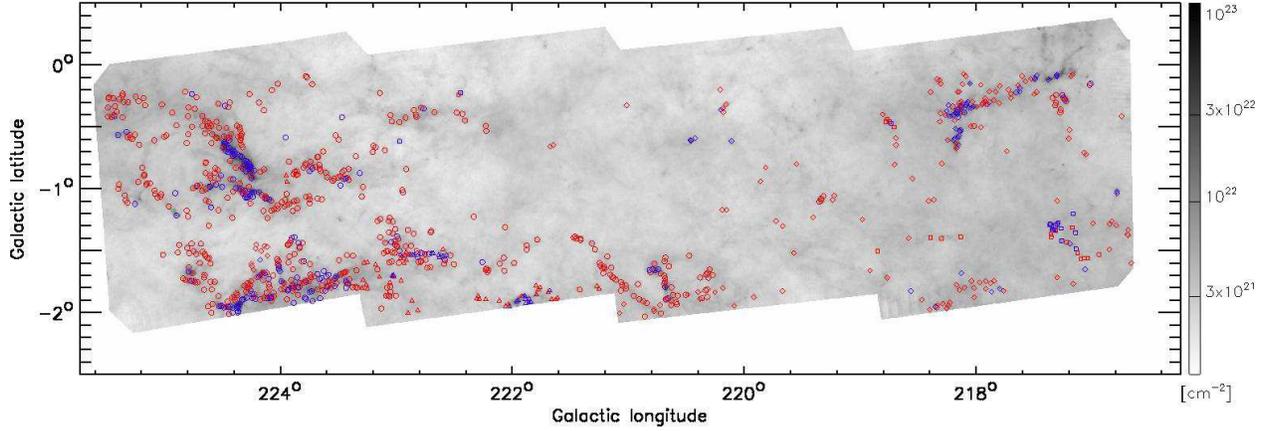}
\caption{Spatial distribution of the Hi-GAL compact sources considered for the analysis in this paper (see text), overplotted on
the column density map of Figure~\ref{hiiregions}. Different symbols are used for sources associated to the four CO components
identified in Section \ref{comaps_par}: circles, triangles, diamonds and squares correspond to the component I, II, III, IV,
respectively. The blue and red colors are used to indicate proto-stellar and starless sources, respectively.\label{sourcepos}}
\end{figure*}

The positions of these sources in the surveyed region are reported in Figure~\ref{sourcepos}, where we introduce the 
color convention blue for proto-stellar objects, and red for the starless ones. We remark that some bright sources might 
have been ruled out from the present statistical analysis by the constraints we applied to deal with regular SEDs.

\begin{figure*}
\includegraphics[width=17cm]{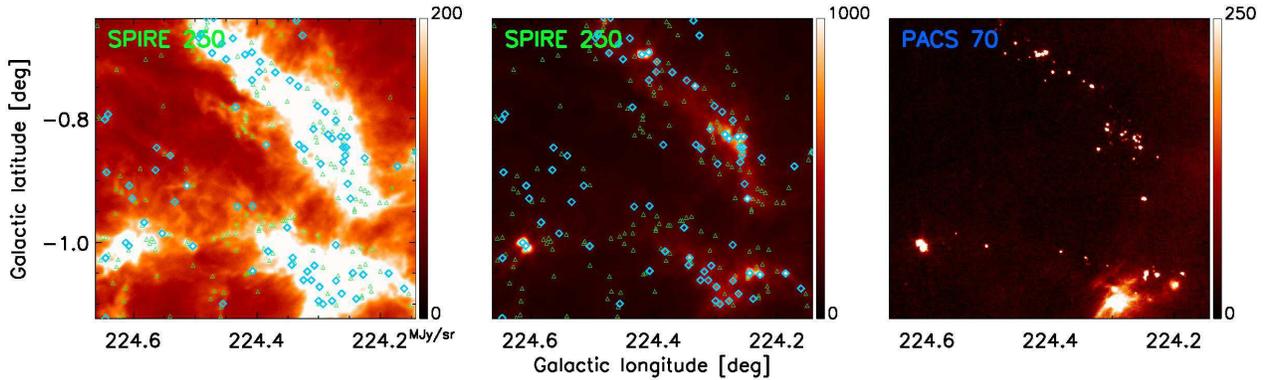}
\caption{Positions of the sources detected in the Hi-GAL 250 $\mu$m map in the neighborhoods of [KKY2004]4.
\textbf{\emph{Left}}: the map is rendered using the same intensity scale as in Figure~\ref{l224}, so that the brightest filaments
appear as white uniform regions in which many compact sources are found (big cyan diamonds: sources admitted in
the final sample; green small triangles: detections rejected for subsequent analysis). \emph{Center}: to appreciate
how these sources actually correspond to local intensity peaks also along the filaments, the same map is rendered
using a less compressed intensity scale. \emph{Right}: the Hi-GAL 70 $\mu$m map, in which both diffuse emission and filaments
are typically much fainter than the sources, is shown to highlight the correspondence of the on-filament sources across the
different bands, as an indication of their genuineness.
\label{extract250}}
\end{figure*}

Furthermore we draw attention on the fact that many sources appear to lie on the most prominent filaments present in the maps. 
Although these filaments appear uniform in figures like those shown in Appendix~\ref{allmaps} due to the choice of the intensity 
scale, in actual fact they possess a complex internal structure revealed by Herschel and characterized by the presence of 
numerous compact sources \citep[see, e.g., ][]{mol10b}. This aspect is well exemplified by the Figure \ref{extract250},
where the positions of the sources extracted at 250 $\mu$m in the neighborhoods of the [KKY2004]4 cloud are superimposed on
the map of this wavelength rendered by means of two different intensity scales. The first one is appropriate to show the 
appearance of the diffuse emission, while the second one is needed to reveal the internal structure of the bright filaments, 
where it is possible to find compact sources whose genuineness is confimred, for example, by the comparison with the 
70 $\mu$m map. \textbf{The occurrence of false positives in our extraction can not be excluded a priori. However, since the detection
is performed independently at each waveband, it is unlikely to systematically find counterparts of them at the other wavelengths. In any
case the source selection described in this section, based on the regularity of the collected SED, ensures the reliability
of the sources considered for the subsequent analysis.}

\subsection{Distance assignment}\label{hdistance}
The problem of the distance estimation of the Hi-GAL sources is a crucial point that has to be faced to derive a 
complete and meaningful set of physical properties for these objects. Fortunately in the portion of the Galactic
plane hosting the $\ell217-224$ field this issue turns out to be less problematic, thanks to both the low degree of 
confusion along the line of sight and the absence of the near/far ambiguity on the kinematic distance, as pointed out 
in Section~\ref{comaps_par}.

A possible approach might consist in examining the line of sight in the CO(1-0) data cube corresponding to an Hi-GAL source, 
and to take the $v_{lsr}$ of the brightest component encountered, as done for example in
\citet{rus11}. Here however we want to exploit the additional information offered by the CO(1-0) data cube decomposition in
clumps (Section~\ref{cdecomp}): in the CO(1-0) line of sight containing the source, we consider the spectral
channels assigned by CLOUDPROPS/CLUMPFIND to one or more clumps, and associate to the source the distance of the clump,
if any, having the closest centroid. There is a double advantage in adopting this strategy: first, the distance assigned
to all the sources associated to a given clump is the same, without finding a different distance at each spatial/spectral
position in the CO(1-0) cube, perhaps affected by the clump internal kinematics unrelated with Galactic rotation. Second, this
strategy acts quite efficiently in cases of presence, along the same line of sight, of a peripheral CO(1-0) feature of a large
and bright clump and of a weaker but central feature of a smaller clump: in this case we believe that the Herschel source
is more likely associated with the small clump being centered on it, rather than with the outer part of a large clump.
Taking into account only the brightest line in this example, instead, would have lead to a completely different association.

\begin{figure}
\epsscale{1.1}
\plotone{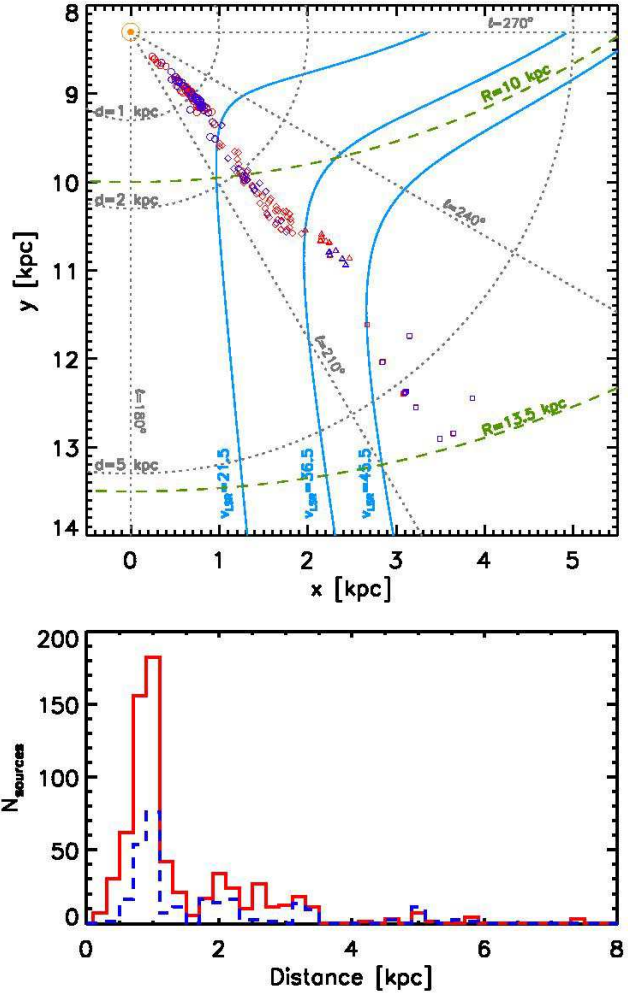}
\caption{Top: sketch of the the TGQ containing the positions of the $\ell217-224$ sources selected for the SED fit. The origin of
the axes is in the Galactic center, and $x$ and $y$ designate the components of the Galactocentric distance (with $x$ and $y$
increasing towards the direction of $\ell=270^{\circ}$ and $\ell=180^{\circ}$, respectively. The symbol and color conventions
are the same of Figure~\ref{sourcepos}. The orange $\odot$ symbol represents the Galactic position of the Sun. The grey dotted
lines indicate relevant longitude values, reported in the labels of the same color. The grey dotted and green dashed arcs
represent relevant values of heliocentric and Galactocentric distance, respectively. Finally, the cyan solid lines are isotachs
corresponding to the three $v_{lsr}$ values separating the four main CO(1-0) components we identified in this paper for convenience.
Bottom: histograms of the source heliocentric distances of the sources shown in the upper panel (blue for 
protostellar and red for starless, respectively). The bin width is of 200 pc.
\label{galpos}}
\end{figure}

The distance information can be exploited to generate a view of the spatial distribution of the sources in the Galactic plane
(Figure~\ref{galpos}, upper panel). In this figure we notice a certain degree of continuity in the distribution 
of the heliocentric distances, without significant gaps between the components I and II. To better illustrate how this 
classification works, the isotachs corresponding to the velocities conventionally separating the four components are also 
plotted. According to the spiral arm loci from \citet{cor02}\footnote{In this case we cannot invoke the careful and more recent 
spiral arm determinations by \citet{val05}, because they were obtained based on a Galactic rotation curve very different 
($R_0=7.9$~kpc) from the one we adopt here.}, in the longitude range we consider in this paper the Local arm should be 
encountered at $d\sim 0.8$~kpc, the Perseus arm at $d\sim 3$~kpc, and the Outer arm at $d>9$~kpc, respectively. 

Our results, in the limit of the reliability of the adopted rotation curve in this part of the Galaxy, partially 
confirm this view. Looking at the lower panel of Figure \ref{galpos}, 
the highest concentration of component I sources is found in the bins centered at 0.8 and 1~kpc. Instead, the distance
distribution of the component II sources peaks around 2, 2.6, and 3.2~kpc, suggesting instead the presence of an inter-arm 
bridge of matter similar to those found by \citet{car05} in other locations of the TGQ. The 0.7-1.1, 2.0, 2.6 and and 
3.2~kpc distances also correspond with the global distances (photometric or kinematic) assigned by \citet{kim04} to their 
clouds present in our field. The Outer arm is not detected in this portion of the TGQ, in agreement with \citet{car05} and 
\citet{xu09}. Finally, we notice that none of the sources we found at the farthest Galactocentric distances can be 
classified as belonging to the far outer Galaxy \citep[$d> 15$~kpc][]{hey98}.

\subsection{SED fitting}\label{sedfit}
As said above, a modified black body (described by the Equation~\ref{greybody}, but with the terms $N(H_2)\, \Delta\vartheta_{500}^2$
replaced by $M/d^2$) has been fitted to the SEDs, to obtain concomitant estimates of the mass and the temperature for each individual
source.

\begin{figure*}
\epsscale{2.0}
\plotone{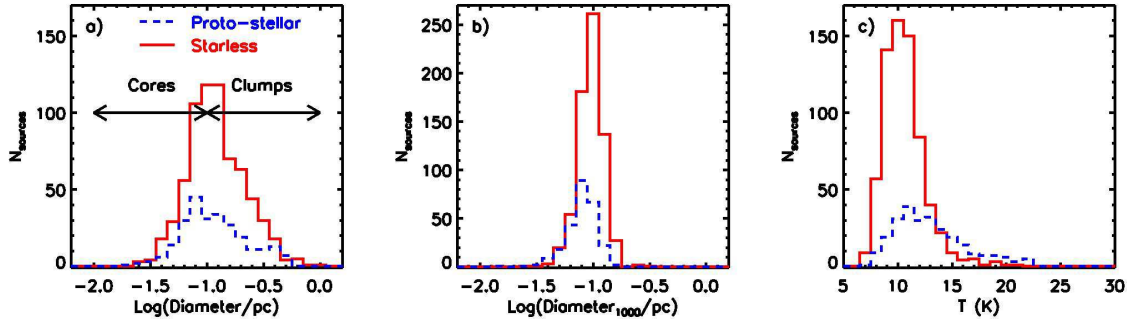}
\caption{Histograms of diameter (panel $a$) and temperature (panel $b$) of starless (red) and proto-stellar
(blue) sources in $\ell217-224$. \label{sizetemp}}
\end{figure*}

In Table \ref{higaltab} the astrometric, photometric and physical properties of the analyzed sources are resumed. Here we start 
discussing the physical diameter and the temperature (Figure~\ref{sizetemp}). 
For sources provided with a distance estimate, the diameter $D$ is easily obtained from the source deconvolved FWHM estimated
at 250 $\mu$m. The histogram in panel $a$ shows that our sample is composed by a mixture of cores ($D<0.1$~pc) and clumps 
($D>0.1$~pc), according to the typical definitions of these object classes based on their size \citep[e.g.,][]{ber07}. 
It is important, therefore, to keep in mind that a relevant fraction of the compact sources analyzed in this work do not 
correspond to progenitors of single star sytems but, rather, to distant complex and large structures we cannot resolve
(in particular, 153 out of 255 proto-stellar sources must be classified as clumps).

The two size distributions do not appear much different each other ($\langle D \rangle = 0.16~\mathrm{pc}$ in both cases), 
unlike in \citet{gia12}, where proto-stellar sources were found 
to be on average smaller that the starless ones. To better discuss this point, we notice that sources spread on a wide range 
of distances are used all together to build the histograms for both the samples, so that some large proto-stellar clumps located 
at long distances are obviously larger than closer pre-stellar cores. For this reason we show also the distributions of 
the diameters as they would appear if placed at a distance of 1000 pc (panel $b$). Also in this case, there is no clear 
indication of significant size differences. In fact, \citet{gia12} adopted a more flexible criterium to decide the reference 
wavelength at which the source size is estimated: 160 $\mu$m when a detection at this wavelength were available, and 
250 $\mu$m otherwise. Since many starless sources are not detected at 160 $\mu$m, their size was estimated at 250 $\mu$m, 
where it appears generally larger (Section \ref{sedbuild}), thus favoring the observed segregation in size, while in our 
case the reference wavelength is the same in all cases.  

Concerning the temperature distributions of the two classes, also in this case the dichotomy proto-stellar-warmer against
starless-colder is not evident as in \citet{gia12}, again due to different reference size adopted in that case implying a 
more consistent flux scaling at the SPIRE wavelengths for the proto-stellar sources, finally leading to higher temperature 
and lower mass estimates. In our case, the uniform choice of the size at 250 $\mu$m as reference reduces the differences of 
the two distributions, that can be found in the presence of a significant high-temperature tail for the proto-stellar sources, 
and in the average temperatures ($\langle T_{ps} \rangle = 13.6~\mathrm{K}, \langle T_{sl} \rangle = 11.1~\mathrm{K}$). It is worthy of notice that the
temperature is estimated on the range 160-500~$\mu$m: this implies that the difference between proto-stellar and starless 
sources is independent from the presence of a flux at 70~$\mu$m, i.e. the constraint we impose to classify the proto-stellar 
sources. Finally note that, unlike the size distribution, in this case there is no distance effect biasing the observed trend 
of the temperature.

\subsection{An estimate of the clump and star formation efficiency}
The combination of Hi-GAL and NANTEN data allow us to provide an estimate of the clump and star formation efficiency (CFE 
and SFE, respectively) for the four distance components seen towards $\ell217-224$. Indicating the total mass of the compact
sources (both proto-stellar and starless) with $M_{cl}$ and the total cloud mass with $M_{gas}$, the CFE is expressed by 
$CFE=\frac{M_{cl}}{M_{cl}+M_{gas}}$. The cloud masses can be easily derived from the column density map of 
Figure~\ref{hiiregions}, resolving the infrequent cases of overlap of two distance components along a CO line 
of sight by adopting the distance corresponding to the brightest feature in the spectrum. Due to the different spatial 
coverage of the Hi-GAL and NANTEN observations, for each component we derive again the cloud mass only in the 
lines of sight falling inside the common area. For the same reason, to compute $M_{cl}$ we must consider only 
the Hi-GAL sources lying in the area surveyed in the CO(1-0) line, but all the sources considered for the present analysis 
fulfill this requirement because they are required to have an associated kinematic distance. Since the 
column density and the compact sources masses have been computed using the same opacity $k_0$, our CFE will not be affected 
by the large uncertainties associated to the choice of this constant. The obtained CFEs for the four components 
are $0.055 \pm 0.008$, $0.024 \pm 0.003$, $0.07 \pm 0.01$, and $0.027 \pm 0.004$, respectively. 
The uncertainties have been estimated assuming a conservative error of 10\% on both the source masses and pixel-to-pixel
column density, combined in quadrature.

In the same way, if $M_*$ is the total mass of the formed stars, the SFE is expressed by $SFE=\frac{M_*}{M_*+M_{gas}}$. 
To compute $M_*$ we consider only the Hi-GAL proto-stellar cores lying in the area surveyed in the CO(1-0) line, obtaining 
core total masses of 2880, 2540, 2140, and 2550~$M_{\odot}$ for the four components, respectively. Finally, these values
have been converted in stellar masses assuming a core-to-star conversion efficiency $\varepsilon_{\mathrm{eff}}=
1/3$ \citep{alv07,and10}, and a average SFE of $8^{+2}_{-1}\times10^{-3}$, $3.9^{+0.9}_{-0.6}\times10^3$, 
$1.2^{+0.3}_{-0.2}\times10^{-2}$, $5.8^{+1.2}_{-0.8}\times10^{-3}$ has been obtained in the four cases. Here the relative uncertainties 
are not simply the same associated to the CFE, since one must take into account also the assumptions on the 
$\varepsilon_{\mathrm{eff}}$. Following \citet{kau10}, we adopt $\varepsilon_{\mathrm{eff}}=1/2$ as upper limit 
for this coefficient.

Further considerations, however, lead us to take these estimates as lower limits for the real SFE:  $i$) this 
derivation of the SFE is based only on FIR/sub-mm observations, neglecting the mass of more evolved young stellar 
objects detected only in the medium infrared; $ii$) as it will be discussed in Section~\ref{evol}, it 
is possible that a small fraction of sources classified as starless cores, not taken into account for this 
calculation, may already have a strongly embedded early-stage proto-stellar content. 

On the other hand, at increasing distances the so-called compact sources
correspond to structures having increasing physical sizes, departing from the definition of core. Therefore,
unresolved distant regions are considered as a single source, although containing in most cases a complex internal
substructure consisting of dense cores and diffuse matter. In particular, for such sources classified as proto-stellar 
clumps, only a fraction of the mass is taking part in the ongoing star formation processes, nevertheless the total
mass is entirely taken into account in the SFE calculation.

In any case, these results are comparable with typical SFE values. In the 40 star forming complexes analyzed by
\citet{mur11}, the SFE ranges from 0.002 to 0.2, with an average of 0.08. In \citet{lad10}, who analyzed 11 nearby
molecular clouds, it ranges from $\sim 0.003$ (Pipe nebula) to $\sim 0.09$ (RCrA). Compared with these data, the 
SFE of our four components can be classified as relatively low. It is noteworthy, in this sense, that we
find the lowest SFE in what we call component II, that is likely spatially related to the Maddalena's 
cloud, i.e. one of the most striking cases of giant molecular with an unusual combination of high gas mass and 
little evidence for star formation \citep[e.g.,][]{hey06,meg09}.

\subsection{Mass distribution of the sources}\label{higalprops}
In the following discussion we focus on the starless objects that are also gravitationally bound, which we call pre-stellar 
cores, because they may give rise to the formation of proto-stellar clusters. In the recent Herschel literature, to study 
the stability of such sources, the lack of kinematic information has forced the adoption of the critical Bonnor-Ebert (BE) 
mass as a surrogate for the virial mass \citep{kon10,gia12}. We use it also in the present case: although in principle we 
could exploit the velocity dispersion obtainable the CO(1-0) data to estimate the virial mass, in practice the CO(1-0)
line spectral resolution is too coarse for determining reliable line widths.

The BE mass is given by:
\begin{equation}\label{mbe}
M_{\mathrm{BE}} = 2.4~r_{\mathrm{BE}}~a^2 /G \;,
\end{equation}
where $a$ is the sound speed at the source temperature, $G$ is the gravitational constant, and $r_{\mathrm{BE}}$ is the 
BE radius (in pc), that in this case is assumed to coincide with the source radius observed at the reference wavelength. 
We consider gravitationally bound, and therefore pre-stellar, all the starless sources with $M> M_{\mathrm{BE}}$
\citep[cf.][]{olm10,gia12}. According to this definition, we find 398, 131, 42 and 19 pre-stellar sources associated to the 
gas velocity component I, II, III, and IV, corresponding to the 82\%, 94\%, 98\% and 100\% of the starless core populations,
respectively.

\begin{figure}[ht]
\epsscale{1.0}
\plotone{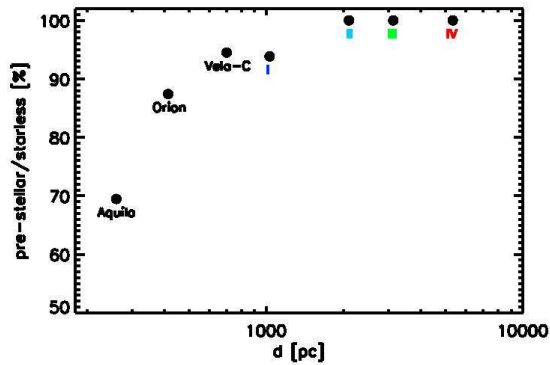}
\caption{Fraction of pre-stellar sources over the total number of starless ones for seven different regions, as a function 
of the average region distance. Data from previous literature have labels in black, while data from this article have 
labels with different colors, according with the color-component convention adopted in Figure~\ref{vfield}. \label{prefrac}}
\end{figure}

In our case the availability of four different distance components gives us the chance to discuss a trend emerging from 
the recent Herschel literature. The studies of the relatively nearby star-forming regions of Aquila \citep{kon10}, Orion-A
(Polychroni et al. 2013, in prep.), and Vela-C \citet{gia12}, located at 260, 414 and 700~ pc away, respectively,
provided an estimate of the pre-stellar fraction over the total starless core number of 69\%, 87\%, and 94\%, respectively. 
These percentages have been estimated using a less severe threshold for identifying the gravitationally bound sources, 
namely $M>0.5 M_{\mathrm{BE}}$. If we adopt the same criterium for the components I-IV of the present work, the fraction 
of pre-stellar sources increases to 94\%, 100\%, 100\%, 100\%, respectively. We notice a general trend of this fraction 
to increase with the distance (see Figure~\ref{prefrac}), highlighting a clear selection effect against gravitationally
unbound structures located at large distances (at $d\geq 2000$ pc the saturation is practically achieved). This is 
implicit in fact in Equation \ref{mbe}: given a certain SED and changing its distance $d$, $M$ varies as $d^2$, while
$r_{\mathrm{BE}}$ varies as $d$, thus making it easier for the same SED to fulfil the BE criterium when $d$ increases. 
On one hand, this means that to exploit this fraction of gravitationally bound sources in a region as an evolutionary 
indication about its star formation capability is affected by a strong and practically unsolvable bias related to the 
distance. On the other hand, the slight deviation of the behavior of the component I from the general trend seen in 
Figure~\ref{prefrac} can be considered, from the qualitative point of view, indicative of an intrinsically lower 
occurrence of pre-stellar sources compared, for example, with the Vela-C region.

\begin{figure}[ht]
\epsscale{1.0}
\plotone{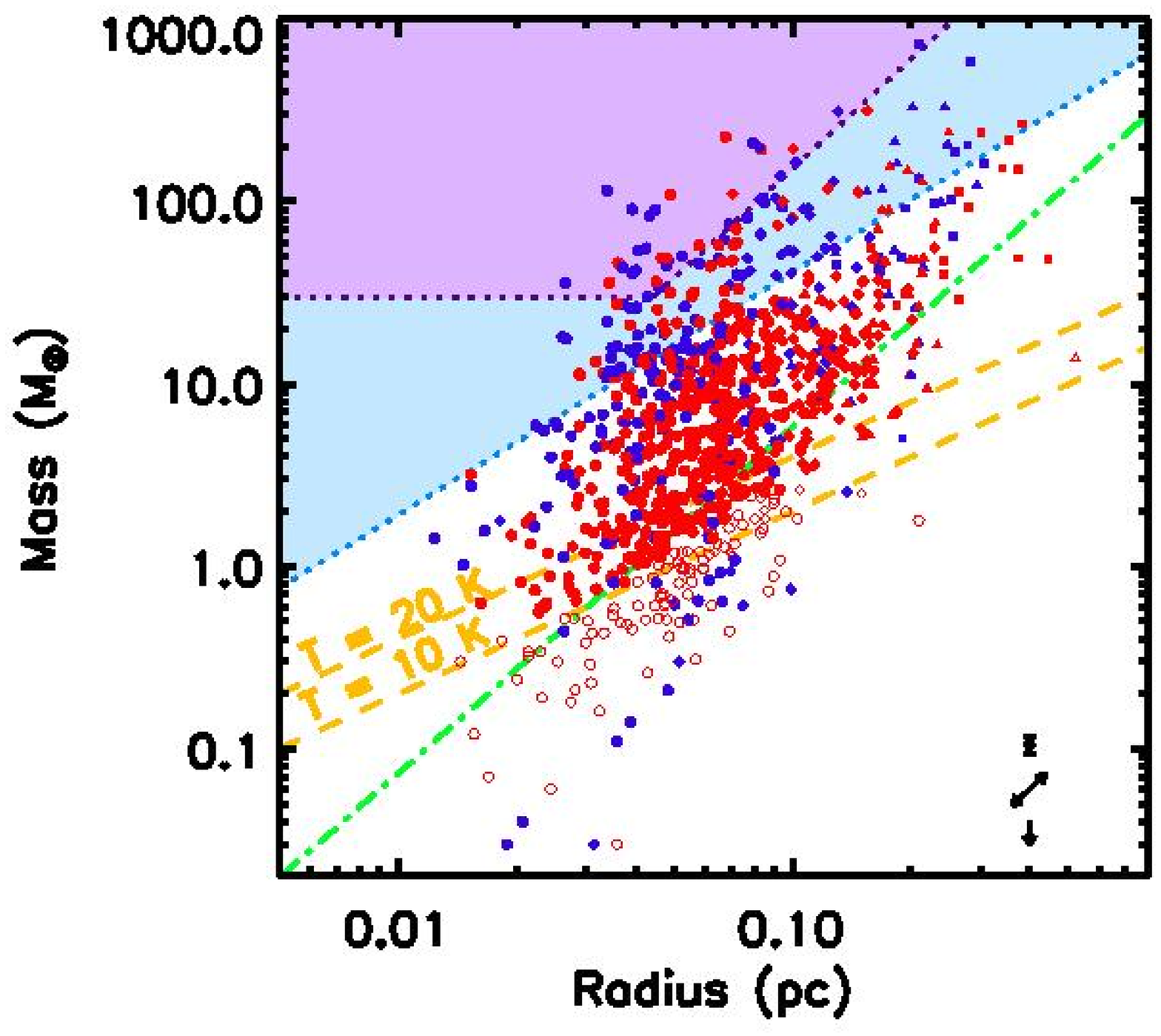}
\caption{Mass vs radius diagram for the Hi-GAL sources in $\ell217-224$. Symbols with the same shape convention of
Figure~\ref{sourcepos} represent pre-stellar (red filled), starless unbound (red open), and proto-stellar (blue
filled) sources. Orange dashed lines are the loci of sources with $M = 1 M_{\mathrm{BE}}$, for $T = 10$ K and
$T = 20$ K, while the green dotted-dashed line represents the power-law estimated by \citet{lar81}. The area with
a purple background represents the region corresponding to the conditions for MSF to occurr, according to the 
\citet{kru08} threshold (purple dotted line). It represents a sub-region of the blue-background area, 
corresponding to the MSF threshold of \citet{kau10} (blue dotted line). In the bottom right corner of 
the plot three possible sources of uncertainty are represented: the error bar represents a 10\% uncertainty on the mass 
coming from the original photometry, the double arrow indicates the shift that would be produced by a $\pm 10\%$ uncertainty 
on the distance, and the down arrow indicates the shift that would be applied if a different opacity law was
adopted, as proposed in Section \ref{conh2}.
\label{masssize}}
\end{figure}

A mass vs radius plot, shown in Figure~\ref{masssize}, can be used to illustrate and quantify the discussion
on the gravitational stability of the starless cores (neglecting the turbulence and magnetic field supports against 
gravity). The lines corresponding to 1~$M_{\mathrm{BE}}$ mass behavior at two different temperatures, $T=10$~K 
and 20~K, respectively, are displayed (orange dashed lines). It can be seen that all the unbound sources (red 
open symbols) lie below the curve at $T=20$ K, while, at the same time, most of the bound ones (red filled symbols) 
show a remarkable degree of boundedness (i.e. $M \gg M_{\mathrm{BE}}(T=20 \mathrm{K})$). Although this diagram is 
meaningless in the case of the proto-stellar sources, since part of the mass has already formed the central 
protostar(s) or could have been ejected in form of proto-stellar jets, we report also these sources in the plot 
(blue filled symbols) for completeness.

In Figure~\ref{masssize} ``Larson's third law'' is shown as a green dotted-dashed line. Originally formulated as
$M(r)>460 M_{\odot} (r/\mathrm{pc})^{1.9}$ \citet{lar81} it has been revised later: \citet{sol87} found a value 2 for
the power-law exponent, i.e. a constant column density as a function of the radius. Since the two values are very similar,
we plot this relation in its original form. In any case, it can be seen that the sources of our sample do not obey a
precise power law. Therefore we do not try here to fit a power law to the mass-size distribution of the pre-stellar cores
in $\ell217-224$.

The mass vs radius diagram is also a useful tool for checking the presence of the conditions for MSF, based on the intuitive
concept that to form massive stars a conspicuous mass reservoir concentrated in a relatively small volume is required. In the
recent literature a theoretical column density threshold of 1~g~cm$^{-2}$ \citep{kru08} is largely invoked and used to identify
candidate sites of MSF (see also Section~\ref{nh2_sect}). It can be represented as a locus in the bi-logarithmic plot of 
Figure~\ref{masssize}, down to a minimum core mass value that, for a 10~$M_{\odot}$ star and a core-to-star conversion
efficiency $\varepsilon_{\mathrm{eff}}=1/3$, amounts to 30~$M_{\odot}$. A small but significant fraction of sources (both
pre- and proto-stellar) exceed this threshold and can be considered therefore potential progenitors of massive stars. Also, 
an empirical and less demanding limit for MSF has been estimated by \citet{kau10}, based on observations of Infrared Dark 
Clouds \citep[IRDCs][]{ega98}, as $M(r)>870 M_{\odot} (r/\mathrm{pc})^{1.33}$ and reported in Figure~\ref{masssize}. Further 
sources, belonging to all the four distance components of $\ell217-224$, fulfil this threshold, testifying the occurrence 
of the conditions for having MSF also in this portion of the TGQ and, in particular, at remarkably large Galactocentric 
distances ($R>11$~kpc for the component IV).

Moving to the analysis of the mass distribution, it becomes necessary to deal with homogeneous samples, from both the spatial and
the evolutionary point of view. Thus, we consider only the pre-stellar sources of the closest velocity component (I), a choice that
also ensures enough statistics to obtain a reliable mass distribution. Again, it is correct to calculate this kind of distribution
only for the pre-stellar sources, as the masses of the proto-stellar ones are not representative of the initial core mass.

\begin{figure}[ht]
\epsscale{1.05}
\plotone{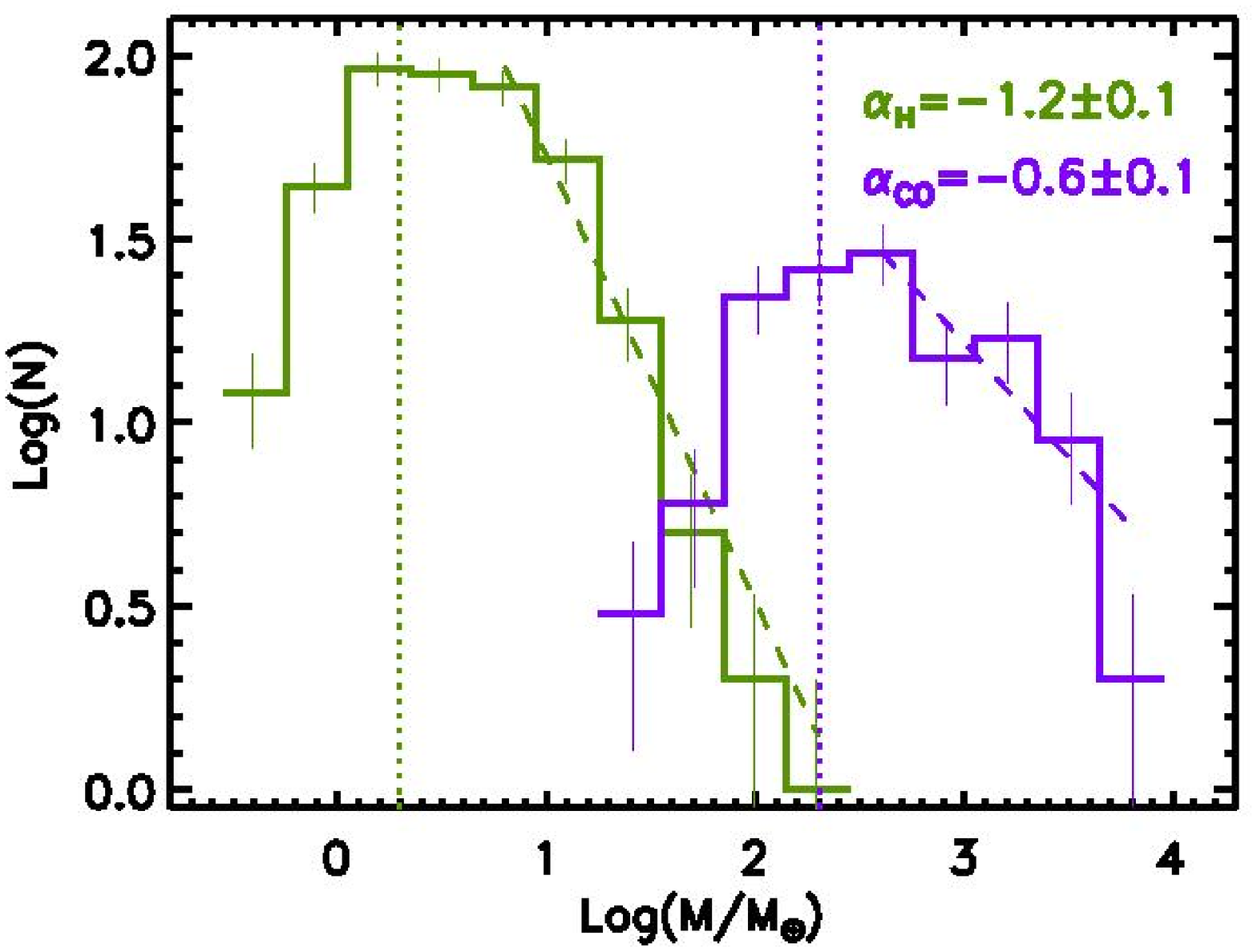}
\caption{Source mass distribution in $\ell217-224$ (green: Hi-GAL pre-stellar sources, purple: CO clumps). The error 
bars are estimated as $\sqrt{N}$ statistical uncertainties. The dashed lines represent the maximum likelihood fit 
\citep[see][]{olm13} to the linear portion of the two mass distributions, and the corresponding slopes are reported. 
As a consequence of the adopted fitting technique, the slope does not depend on the arbitrary choice of the bin
width. The dotted vertical lines represent the mass completeness limit in the two cases: for the Hi-GAL observations it 
has been derived as described in the text, whereas for the NANTEN observations we followed the method of \citet{sim01}. 
\label{cmf}}
\end{figure}

The core mass distribution is reported in Figure~\ref{cmf}. The mass completeness limit of $1.9 M_{\odot}$ has been 
estimated from the flux completeness limit at the reference wavelength 250~$\mu$m (Section~\ref{cores}) using the modified 
black body equation, combined with the average temperature and distance of the sample, and $\beta=2$. The turn over of 
the mass function falls around $3~M_{\odot}$, too close to the completeness limit to claim that we are well tracing the 
log-normal behavior of the mass distribution below the turnover point \citep[as, e.g., in][]{kon10}. 
We then fit a power law $N(\log M) \propto M^{\alpha}$ to the high-mass part of the distribution using an approach 
based on the method of maximum likelihood and independent from the histogram binning \citet[see][and references therein]{olm13}. 
It allows us to determine the lower limit of range of validity of the power-law, $ \log (M_\mathrm{inf}/M_{\odot})=0.65$, and 
its slope $\alpha=-1.0 \pm 0.2$.  

Whereas the core mass functions derived for nearby regions through spectral lines \citep[Orion A, $\alpha=-1.3 \pm 0.1$, ][]{ike07}, 
Spitzer \citep[Perseus, Serpens and Ophiucus, $\alpha=-1.3 \pm 0.2$, ][]{eno08}, and Herschel observations \citep[Aquila Rift, 
$\alpha=-1.45 \pm 0.2$, ][]{kon10} have slope values consistent with those of the stellar initial mass function 
($\alpha=-1.3 \pm 0.7 $ for single stars, 
\citealt{kro01}; $\alpha=-1.35 \pm 0.3$ for multiple systems, \citealt{cha05}), when the considered sample is contaminated 
by the presence of larger structures (clumps) as in our case (see Figure~\ref{masssize}), the slope generally tends to be 
shallower, towards the values found for the clump mass functions (e.g., $\alpha=-(0.6-0.8)$ for various clouds, \citealt{kra98}; 
$\alpha=-0.6$ for the Galactic center region, \citealt{miy00}; $\alpha=-1.0$ for the Vela-D cloud, \citealt{eli07}). A very 
similar behavior was observed also in the case of the Herschel-based source mass distribution of the Vela-C cloud 
\citep[$\alpha=-1.1 \pm 0.2$, ][]{gia12}, affected by the same kind of contamination.

In our case, we can check on the same region this gradual steepening of the mass function from clumps to cores, by considering the
masses of the CO clumps extracted as described in Section~\ref{cdecomp}, and belonging to the component I (purple histogram 
in Figure~\ref{cmf}, $\alpha=-0.7 \pm 0.3$). This steepening is quite usual and is explained, in the general case, as a consequence
of the fragmentation of the clumps in cores that constitute the last fragment generating the single star (or multiple system).
However, in our case, notice that the error bars associated to the two slopes (obtained with the same method) partially overlap. 
Furthermore, the disagreement between Herschel- and CO-derived column densities highlighted in Figure~\ref{nh2co} surely introduces an 
uncertainty on the slope of the CO clump mass function larger than that derived simply from the power-law fit.

It is possible to make a direct comparison with the mass function slope derived by \citet{kim04} for their "Group I" clumps
(namely sources associated with CMa OB1 and G220.8-1.7), $\alpha=-0.59 \pm 0.32$, that is consistent with our result.

\section{Evolutionary framework}\label{evol}
To further investigate the differences between the proto- and pre-stellar sources considered so far, we 
discuss here a plot of the bolometric luminosities vs the envelope masses. In the previous literature
\citep{sar96,and00,mol08} such a diagram has been already used as a meaningful method to infer the evolutionary
status of the compact sources, through the comparison between their positions in the plot and the theoretical
tracks obtained from models of accreting cores. Such tracks start from a given initial mass and initially follow
an almost vertical path corresponding to the accretion phase. Subsequently, the proto-stellar outflow activity
produces mass loss and disperses the residual envelope (horizontal portion of the tracks).

Recently, this diagnostics has been applied to the clump populations of star forming regions observed by Herschel
\citep{eli10,bon10,hen10,gia12,ven13}, revealing some degree of segregation between proto- and pre-stellar sources.
A general caveat about the use of this diagram for distant star forming regions consists in the fact that,
rigorously speaking, the evolutionary tracks are calculated for a single protostar. Instead, in cases like ours,
many sources are in fact clumps having an unresolved internal structure, with the possible presence of few
proto-stars, and/or smaller pre-stellar cores. Therefore, a precise star formation timeline cannot be deduced
from such tracks. Of course also in these cases to find a higher luminosity at the same envelope mass is indicative
of a more evolved evolutionary stage, but it is difficult to understand whether this is a general property of
the entire (unresolved) population hosted by the clump, or it is essentially due to only the brightmost contained
source(s).

To build this diagram we used the mass values discussed in the previous sections. They represent, in fact, 
the envelope mass in the case of a proto-stellar source, and of course the whole mass of a starless one. 
The luminosities have been derived integrating the whole observed SED, including $i$) the flux
of the best-fitting modified black body at 1 mm, to take into account also the millimetric portion of the SED 
we do not observe directly, and $ii$) the WISE flux at 22~$\mu$m, where available, to ensure a 
better estimate in the case of the sources most evolved and then bright in the mid-infrared unlike, for 
example, the similar plot in \citet{gia12}. 
By the way, in \citet{mol08} and \citet{hen10}, since their targets were constituted by limited amounts of 
proto-stellar sources (this explains the lack of the pre-stellar ones in their analysis), more extended SEDs were 
collected, down to 8.3 and 3.6 $\mu$m, respectively. On one hand, this increases and improves the bolometric luminosity 
estimation for more evoled sources, on the other hand this operation requires special care in checking the 
reliability of the short-wavelength counterpart assignment to each single FIR source, and therefore it lies 
outside the goals of the present work.

\begin{figure*}
\epsscale{1.6}
\plotone{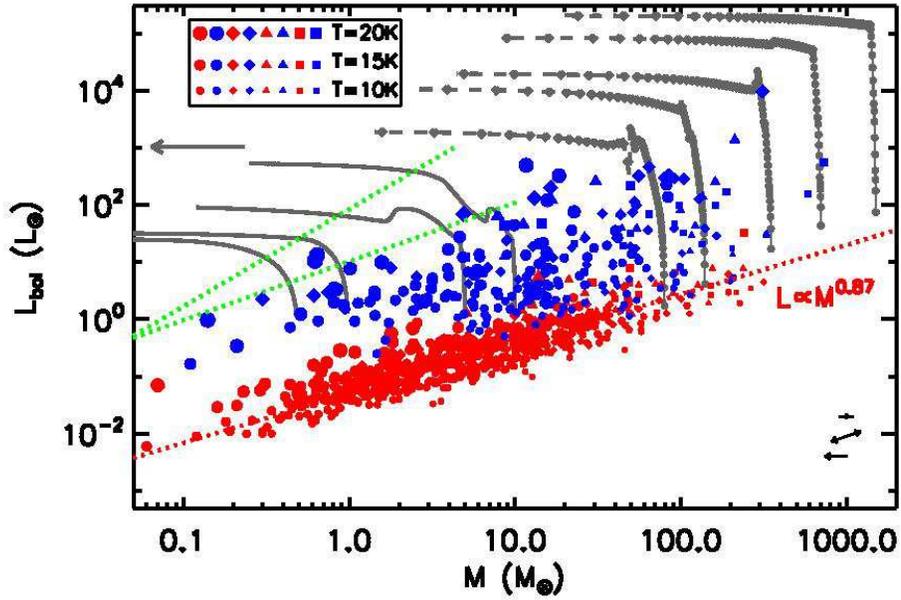}
\caption{Plot of the bolometric luminosity vs the mass of the the same sources of Figure~\ref{masssize}, with the same symbol and color
convention. In addition, the symbol size varies according to the source temperature, as illustrated in the legend. Grey and black solid
lines represent the evolutionary tracks for the low- and high-mass regime, respectively, taken from \citet{mol08}. The grey arrow indicates
the evolution direction, while green dashed lines delimit the region of transition between Class 0 and Class I sources \citep{and00} in the 
low-mass regime. The red dotted line represents the best-fitting power law ($L_{\mathrm{bol}} \propto M_{\mathrm{env}}^{ 0.87 \pm 0.02}$) 
for the distribution of the prestellar sources. In the right bottom corner the effects of possible sources of random or systematic 
uncertainty affecting the points of the plot are represented as arrows, as in Figure \ref{masssize} (see it for further 
explanation).
\label{lm_temp}}
\end{figure*}

The $L_{\mathrm{bol}}$ vs $M_{\mathrm{env}}$ plot is reported 
in Figure~\ref{lm_temp}, whith the same symbol convention of Figure \ref{masssize}. Furthermore, the symbol size 
linearly scales with the source temperature. The theoretical evolutionary tracks for the low and high mass regimes 
adopted by \citep{mol08} are also plotted. 

Also in our case, the proto-stellar sources and the pre-stellar ones populate quite different regions of the diagram,
corresponding to the accreting core phase in the first case, and to a quiescent or collapsing core in the second,
respectively. Obviously, there is a transition region with some overlap between the two populations, especially at high 
masses (discussed below), but a global look suggests a certain degree of segregation. 

In the low-mass regime, where the distinction of well-defined evolutionary classes makes more sense 
\citep[][and references therein]{and00}, the region corresponding to the Class 0 to I transition appears populated, as 
in \citet{gia12}. However, we remind the reader that the $L_{\mathrm{bol}}$ vs $M_{\mathrm{env}}$ plot is not
a tool \textbf{suitable} for obtaining a precise classification, essentially because $i$) \citet{hen10} showed that sources classified as 
Class I from their NIR-to-MIR SED are found to be compatible instead with a Class 0 status based on their 
FIR fluxes, and there is not a sharp transition from Class 0 to I to II in their $L_{\mathrm{bol}}$ vs 
$M_{\mathrm{env}}$ diagram; $ii$) this evolutionary classification can not be easily extended to the high-mass
regime, a fortiori when the considered source is a large clump hosting multiple star formation. 
Furthermore, comparing our diagram with those of \citet{mol08} and \citet{hen10}, we can explain the lack of
evolved sources in ours with the aforementioned different ranges used for calculating $L_{\mathrm{bol}}$.
In summary, due to both the investigated spectral range and mass regime, this $L_{\mathrm{bol}}$ vs 
$M_{\mathrm{env}}$ plot can not be used to obtain a classical single-YSO classification.

From the point of view of the distance, it is remarkable that proto-stellar activity is detected also in the most 
distant sources, confirming the occurrence of the star formation also at $R>11$~kpc \citep[see, e.g., ][and 
references therein]{rei08,yun09}.

Focusing on the source temperatures, the first general indication is that the most luminous proto-stellar sources, i. e. the most
evolved, are also the warmest, and of course the proto-stellar are on average warmer than the pre-stellar ones as already indicated
by Figure~\ref{sizetemp}. Furthermore, the luminous pre-stellar sources occupying regions of the diagram where early-stage proto-stellar
cores are instead expected (beginning of the evolutionary tracks) are generally characterized by a higher temperature. We interpret this
fact as the presence of a strongly embedded early-phase protostar(s) not detected at both the 22 and 70~$\mu$m wavebands. This might lead
to misclassify early proto-stellar sources as pre-stellar, a fact that should be taken into account in the future analyses of core/clump
populations extracted from the Herschel maps.

Finally, we notice that while the proto-stellar sources are spread over a wide range of possible evolutionary stages, the 
displacement of the bulk of the pre-stellar sources follows a power-law in a region of the diagram corresponding to a 
stage in which no mass accretion has started yet. A similar behavior was already discussed in \citet{gia12} for the Vela-C 
cloud but here it looks more evident and, more interestingly, it appears the same over several spatially unrelated regions.
A power-law fit gives a dependence $L_{\mathrm{bol}} \propto M^{(0.87 \pm 0.02)}$. This slope is very similar to the 0.85 value
found by \citet{bra01} for their sample of proto-stellar sources, while it is very different from the 0.42 value found by 
\citet{gia12} for pre-stellar sources in Vela-C. However, we need to increase the statistics of this slope across 
the Galactic plane, exploiting the whole Hi-GAL data base, before stating a possible link between the evolutionary stage of 
a star forming region and the slope of the luminosity vs mass relation of the pre-stellar cores.

\subsection{Star formation rate}\label{sfr}
The $L_{\mathrm{bol}}$ vs $M_{\mathrm{env}}$ diagram can be exploited to quantify the star formation rate (SFR) of the 
four distance components we identify in $\ell 217-224$. For the high mass regime ($M > 10 M_{\odot})$ we followed the method 
of \citet{ven13}. Briefly, it consists in associating each proto-stellar source mass to a nearby evolutionary track, 
and to consider the mass $M_f$ of the central object at the end of the track (corresponding to a time $t_f$), when the 
envelope clean-up phase is completed. The SFR is calculated by adding the $M_f/t_f$ ratios of all sources. In the low-mass regime, 
instead, we use the Equation 1 of \citet{lad10} that assumes a constant final mass and time, $0.5 M_{\odot}$ and $2\times10^6$ yr, 
respectively. Anyway, the contribution of the low-mass part to the total SFR amounts to  8\% for the component I and less than 1\% for
the remaining components. We find, for the components I-IV, $SFR=2.6^{+0.2}_{-0.3}\times10^{-4}, 1.39^{+0.03}_{-0.05}\times10^{-4},
6.30^{+0.04}_{-0.23}\times10^{-5}, 4.77^{+0.26}_{-0.05}\times10^{-5} M_{\odot}$ yr$^{-1}$, respectively. In the portion of 
the Galaxy corresponding to $\ell 217-224$, thus, $SFR_\mathrm{tot}=5.1^{+0.2}_{-0.3}\times10^{-4} M_{\odot}$ yr$^{-1}$.
The error bars have been estimated assuming a 20\% average uncertainty on both mass and luminosity. Of course, larger 
uncertainties come from the choice of the YSO model and from the distance estimate and are not easily quantifiable 
\citep[see][]{ven13}.

In two single Hi-GAL tiles of the first Galactic quadrant, namely centered close to 
$[\ell,b]=[30^{\circ},0^{\circ}]$ and $[59^{\circ},0^{\circ}]$, \citet{ven13} obtained $SFR=9.5\times10^{-4}$ and 
$1.6\times10^{-4} M_{\odot}$ yr$^{-1}$, respectively, considering all the distance components together. Taking 
into account that the solid angle spanned by the $\ell 217-224$ maps is about five times larger than in the 
case of these two inner Galaxy fields, we conclude that in our case the SFR per solid angle is significantly 
lower, as expected for the outer Galaxy. However, the forthcoming studies based on the Hi-GAL survey will
provide a more exhaustive description of the SFR as a function of the Galactic position, and an extrapolation
to the SFR of the entire Galaxy.

\subsection{Diagnostics of CO clumps}\label{lm_co}
If one wants to to apply a diagnostics based on a luminosity-mass relation also to the clumps extracted from the
CO(1-0) data, the most meaningful way is to use the virial mass. Indeed, the way the clump mass is calculated in our case, namely
as proportional to the intensity integrated over all the channels assigned by the cloud decomposing algorithm to the clump
(Equation~\ref{xco}), makes it directly proportional to the luminosity, which strictly speaking is a ``luminous mass'':
\begin{equation}\label{mlum}
M_{CO}=c L_{CO} \qquad .
\end{equation}
\citet{bol08}, who used the same decomposition algorithm we used, report a value for the constant $c=45$ with $M_{CO}$ in $M_{\odot}$
and $L_{CO}$ in K~km~s$^{-1}$~pc$^2$. Instead, the estimate of the virial mass is independent from the clump luminosity. We show the plot
of these two quantities (derived as in Section~\ref{cdecomp}) in Figure~\ref{lm_co}, panel $a$, where we adopt the same symbol convention 
of Figure~\ref{lm_temp} to separate sources associated to the different distance components. Furthermore, blue and red colors are used for 
clumps associated with at least one Hi-GAL proto-stellar source and clumps associated only with Hi-GAL starless ones, respectively. Finally, 
smaller green symbols represent the CO clumps that do not contain Hi-GAL sources, or that lie outside the area surveyed by Herschel.

\begin{figure*}
\epsscale{1.5}
\plotone{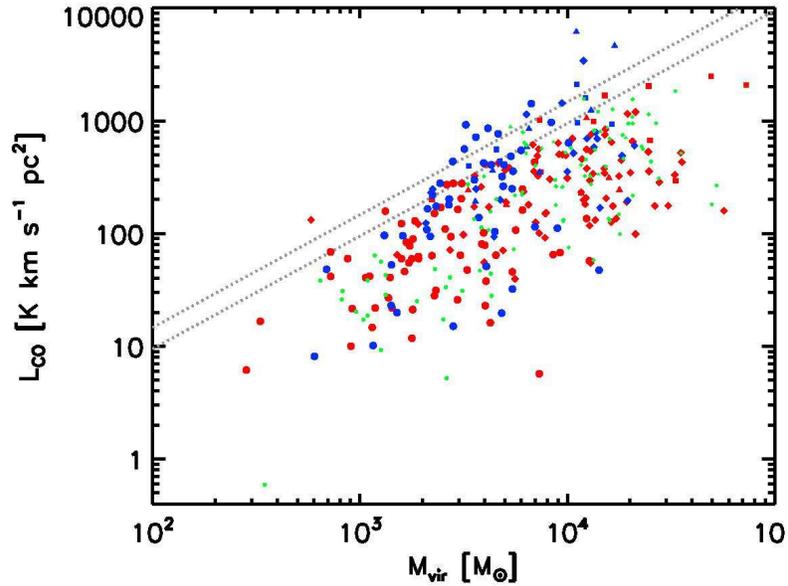}
\caption{Plot of the luminosity vs the virial mass of the CO clumps, derived as in Section~\ref{cdecomp}. The symbol-distance convention 
is the same of Figure~\ref{masssize}. The blue color indicates clumps containing proto-stellar Hi-GAL sources (at least one), while the 
red color indicates clumps associated only with starless Hi-GAL sources. Sources lacking Hi-GAL counterparts or lying outside the area 
surveyd by Herschel are plotted with small green symbols. Two grey dotted lines represent the luminosity vs luminous mass relation for 
the two extreme values of the $X$ factor used in this paper. \label{lm_co}}
\end{figure*}

Two lines corresponding to Equation \ref{mlum} for two different slopes have been plotted (grey dotted lines). They are obtained correcting
the $c$ value of \citet{bol08} (corresponding to $X=2 \times 10^{20}$ (K~km~s$^{-1}$)$^{-1}$ cm$^{-2}$), by the minimum (upper line) and
maximum (lower line) values of $X$ used for the clumps of this paper, respectively, according to Equation~\ref{xco}. This is equivalent
to saying that all the values of the luminosity-luminous mass pairs lie in the stripe of the plot delimited by these two lines (and in
particular the majority of them lies close to the upper one, namely that corresponding to the smaller value of $X$).

The less rigorous classification between ``proto-stellar'' and ``starless'' CO clumps probably explains the reason of the absence of a clear
segregation like that found in Figure~\ref{lm_temp} for dust cores and clumps. It is noteworthy, however, that for $M_{CO}\gg 10^3$ M$_{\odot}$
the most luminous clumps $i$) all have a proto-stellar content; $ii$) all are supervirial. If we consider all the supervirial clumps in our sample,
namely 22 objects corresponding to $\sim7\%$ of the total, 17 of them have a proto-stellar content, 3 are starless and 2 are not associable with
Herschel sources. This indication is surely relevant, but not resolutive or sufficient to identify evolved clumps in such a diagram. On one hand,
the majority of the supervirial structures are found to host star formation; on the other hand, many ``proto-stellar'' clumps can appear as
sub-virial, since their global physical properties plotted here cannot describe local conditions that could allow star formation in limited
parts of them. Finally, unlike the case of the Hi-GAL $L_{\mathrm{bol}}$ vs $M$ diagram, we do not recognize any power-law trend either in the
whole sample of clumps or in the single identified classes.

\section{Conclusions}\label{concl}
The first Hi-GAL observations of the TGQ consist of a ``chunk'' of $\sim20.5$~deg$^2$ corresponding to the longitude
range $216.5\degr \leq \ell \leq 225.5\degr $. Here we presented a first study based on both the overall displacement
of the far-infrared emission in this region and on the statistical analysis of the properties of the compact sources
derived from photometric measurements. The NANTEN CO(1-0) observations have added the fundamental information about gas
kinematics. This study confirms the view of the outer Galaxy as an interesting laboratory to investigate star formation
processes, benefitting of the generally simple structure of the velocity field (compared with the inner Galaxy). 
Summarizing, we found that:
\begin{itemize}
  \item In the Hi-GAL images it is possible to recognize the dust emission from all the velocity components in this sky
  region detected in the line observations presented in this and previous papers \citep{kim04}. In particular we
  identify four components (I-IV) located at average heliocentric distances of 1.1, 2.2, 3.3, and 5.8~kpc respectively.
  The overlap among these components is quite infrequent, making clearer their identification.
  \item The large-scale column density and temperature distribution for the cold component of the dust reveal a variety
  of conditions, with star formation sites preferentially distributed along filaments. Interestingly, two of these are
  likely associated with the S296 H\textsc{ii} region, which might be responsible of triggering star formation
  in these filaments.
  \item We obtained a five-band Hi-GAL compact source catalog of this region,
  complemented with the WISE 22~$\mu$m photometry and with the distance information derived from the spatial
  association with the clumps extracted from the CO(1-0) data set. The classification between proto-stellar
  (provided with a detection at 70~$\mu$m and/or at 22~$\mu$m and 160~$\mu$m) and starless sources has been made.
  \item A modified black body has been fitted to the regular and well sampled SEDs (255 proto-stellar, 688
  starless), to derive temperatures and masses. On average, the proto-stellar sources are found to be warmer than the 
  starless ones. Concerning the source size, 
  moreover, we find that our sample is composed by a mixture of clumps and cores, due to the wide involved distance range. 
  Looking at their spatial
  distribution in the Galactic plane, the local and Perseus spiral arm are clearly seen, although a remarkable
  amount of sources with assigned inter-arm distances are also found. The limited range of investigated Galactic longitudes
  does not allow a better recognition of a grand design spiral pattern.
  \item We derived the SFE for the four velocity components (0.008, 0.004, 0.012, 0.006, respectively). These 
  values lead to classify the star formation in these regions as not very efficient. On one hand,
  being only based on Herschel far-infrared photometry, they should be taken as lower limits. On the other hand,
  a systematic overestimate of the SFE with increasing distance has to be taken into account.
  \item Applying a virial analysis to the starless Hi-GAL sources, large fractions of them (ranging from 82\% for the
  component I to 100\% for the component IV) are found to be gravitationally bound, i. e. pre-stellar. We show
  that these percentages must be considered as upper limits, because they are biased by a selection effect
  related to the distance and implicit in the adopted criterium.
  \item In the mass vs radius plot for the pre-stellar sources we do not recognize any clear power-law behavior.
  We use this tool to check the potential ability of clump/cores
  to form massive stars, based on the different thresholds of \citet{kru08} and \citet{kau10}. In both cases
  we find condensations compatible with high-mass star formation in all the distance components.
  \item A power-law $N(\log M) \propto M^{\alpha}$ has been fitted to high-mass portion of the core/clump mass 
  distribution of the pre-stellar sources belonging to the component I. Similarly to the Vela-C case 
  \citep{gia12}, its slope $\alpha=-1.0 \pm 0.2$ is shallower (although still consistent within the 
  errors) than the stellar IMF slope and the similar slopes also found for cores in nearby star forming regions, 
  and steeper than the slopes found for the gas clump distributions, as in this case ($\alpha=-0.7 \pm 0.3$).
  \item The $L_{\mathrm{bol}}$ vs $M_{\mathrm{env}}$ plot of the Hi-GAL sources reveals an evident 
  separation between proto- and pre-stellar sources. For both classes the warmest sources are, in general, 
  the most evolved of their class. Few proto-stellar sources populate
  the region of the diagram where objects in transition between Class~0 and Class I are expected to lie, whereas 
  the majority populate the Class~0 region. This classification obviously weakens when distant sources are 
  considered, namely those corresponding to large and massive ($M \gtrsim 100 M_{\odot}$) clumps.
  Using this diagram as in \citet{ven13} we are able to derive also an estimate of the star formation rate 
  of  $SFR_\mathrm{tot}=5.1\times10^{-4} M_{\odot}$ yr$^{-1}$.
  \item Most of the CO clumps (93\%) are subvirial. Line observations at better angular resolution would probably
  reveal denser supervirial substructures. The remaining 7\% fraction of supervirial clumps (22 objects) is
  in most cases associated with Hi-GAL proto-stellar sources, suggesting that these clumps fulfill globally and
  not only locally the conditions for the gravitational collapse to occurr.

Besides the characterization of star forming regions in the $\ell 217-224$, this article gives also a first contribution
to the larger statistics that will be built exploiting the entire Hi-GAL survey archive. Prescriptions and
caveats on using the photometric data, on combining them with kinematics derived from CO observations, and
on addressing effects of distance-resolution have been explicitly discussed in several points of the text.

\end{itemize}

\acknowledgments
PACS has been developed by a consortium of institutes led by MPE (Germany) and including UVIE (Austria); KU Leuven, CSL, IMEC (Belgium); 
CEA, LAM (France); MPIA (Germany); INAF-IFSI/OAA/OAP/OAT, LENS, SISSA (Italy); IAC (Spain). This development has been supported by the 
funding agencies BMVIT (Austria), ESAPRODEX (Belgium), CEA/CNES (France), DLR (Germany), ASI/INAF (Italy),and CICYT/MCYT (Spain). SPIRE 
has been developed by a consortium of institutes led by Cardiff Univ. (UK) and including: Univ. Lethbridge (Canada); NAOC (China); 
CEA,LAM (France); INAF-IFSI, Univ. Padua (Italy); IAC (Spain); Stockholm Observatory (Sweden); Imperial College London, RAL, UCL-MSSL, 
UKATC, Univ. Sussex (UK); and Caltech, JPL, NHSC, Univ. Colorado (USA). This development has been supported by national funding agencies: 
CSA (Canada); NAOC (China); CEA, CNES, CNRS (France); ASI (Italy); MCINN (Spain); SNSB (Sweden); STFC, UKSA (UK); and NASA (USA). DE, 
MP, and KJLR are funded by an ASI fellowship under contract numbers I/005/11/0 and I/038/08/0. DP is funded through the Operational 
Program ``Education and Lifelong Learning'' and is co-financed by the European Union (European Social Fund) and Greek national funds.

{\it Facilities:} \facility{Herschel Space Observatory}, \facility{NANTEN}.

\appendix

\section{Hi-GAL images: the $\ell224$, $\ell222$, $\ell220$, and $\ell217$ tiles from 70 to 500 $\mu$m}\label{allmaps}

\begin{figure*}
\centering
\includegraphics[width=15.0cm]{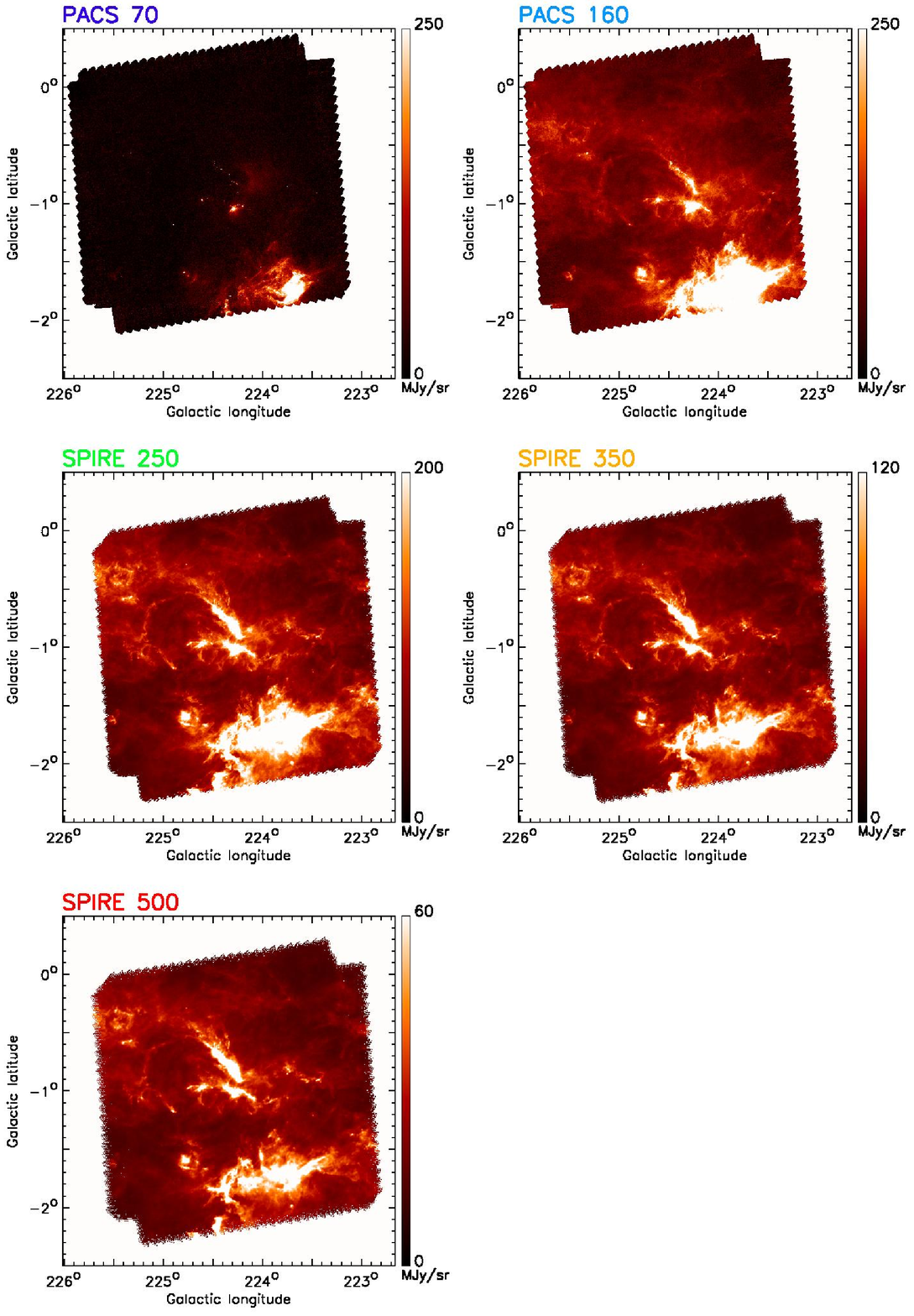}
\caption{Hi-GAL maps of the $\ell224$ field at 70, 160, 250, 350, 500~$\mu$m.\label{l224}}
\end{figure*}
\begin{figure*}
\centering
\includegraphics[width=15.0cm]{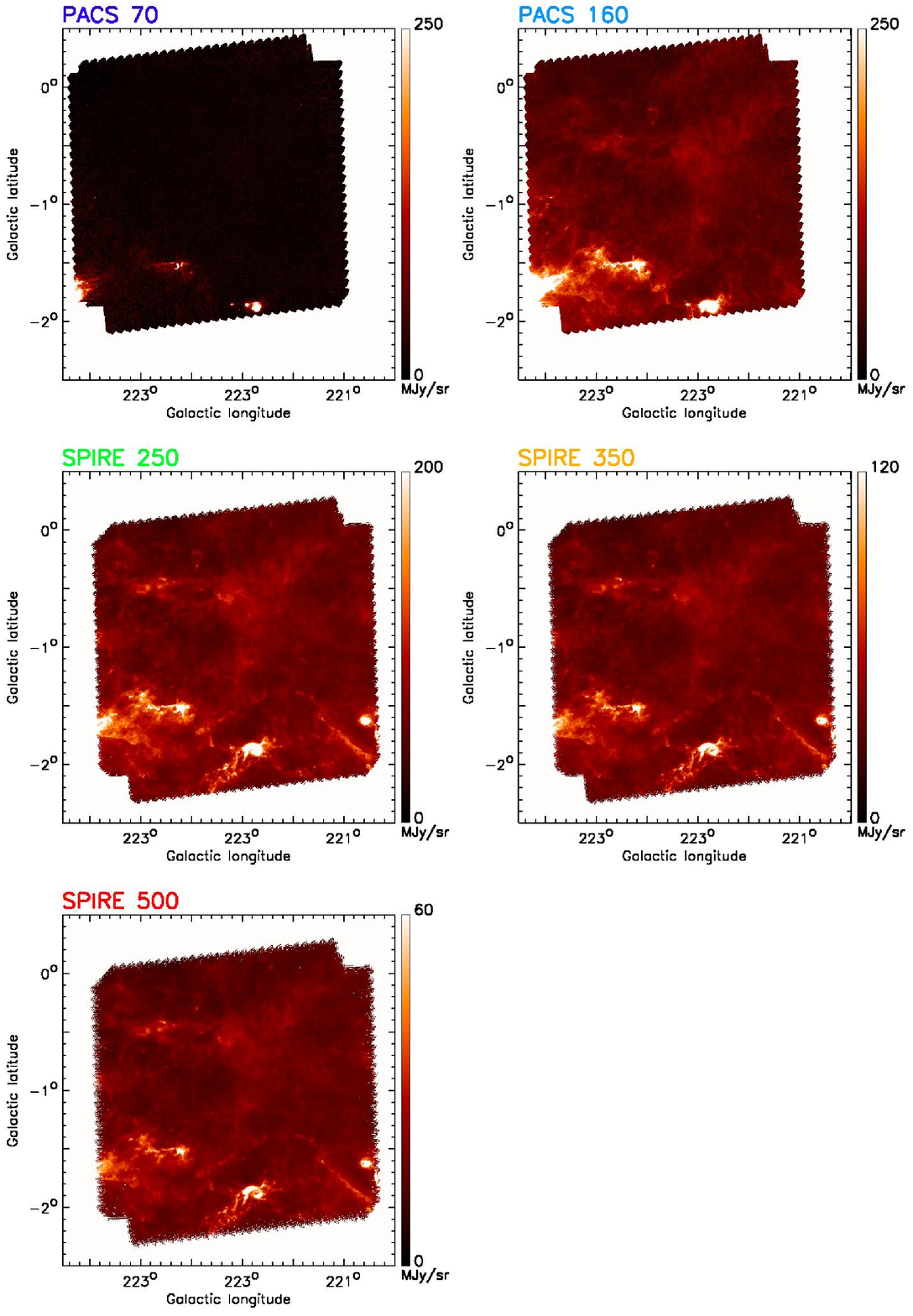}
\caption{The same as Figure \ref{l224}, but for the $\ell222$ field.\label{l222}}
\end{figure*}
\begin{figure*}
\centering
\includegraphics[width=15.0cm]{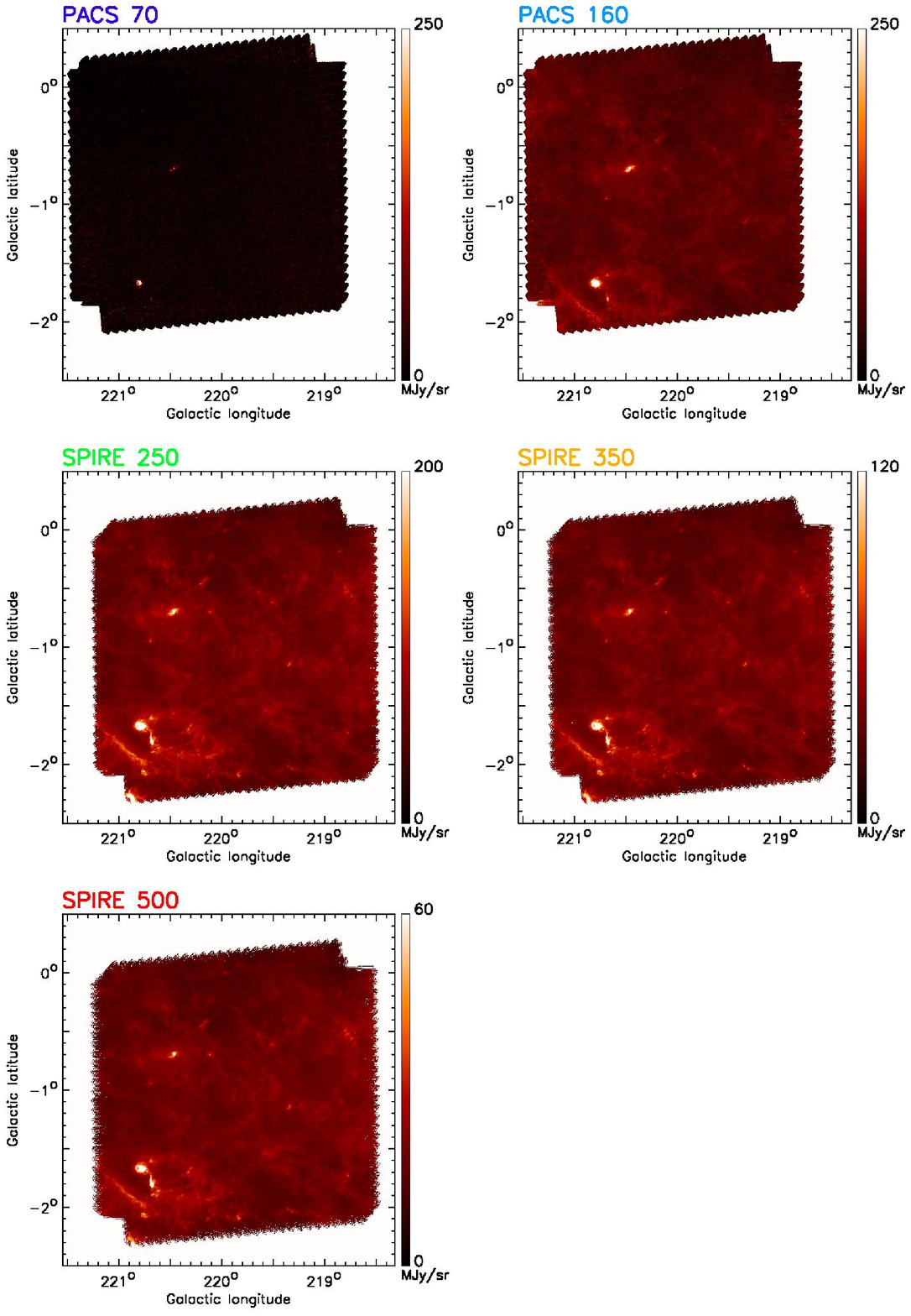}
\caption{The same as Figure \ref{l224}, but for the $\ell220$ field.\label{l220}}
\end{figure*}
\begin{figure*}
\centering
\includegraphics[width=15.0cm]{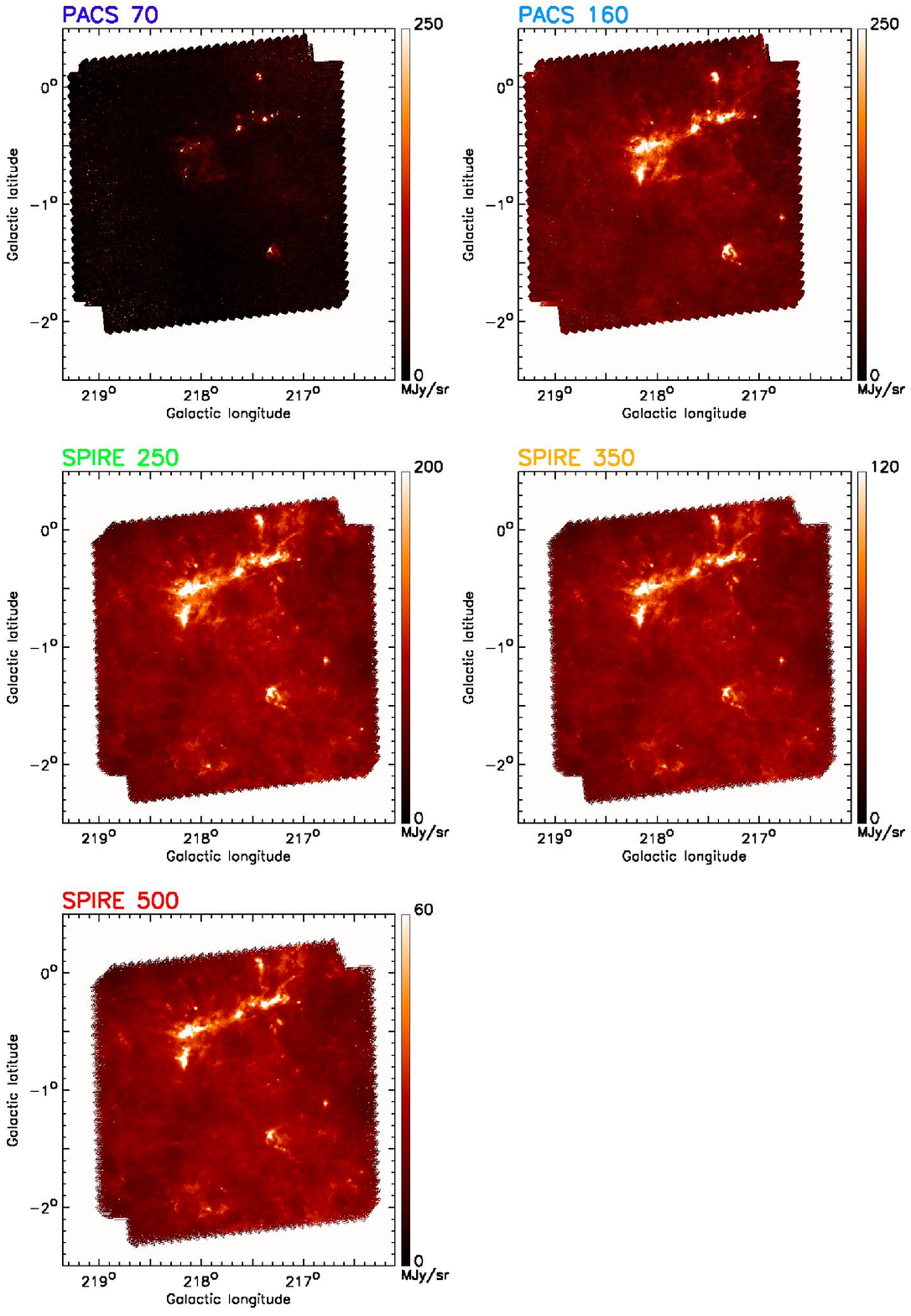}
\caption{The same as Figure \ref{l224}, but for the $\ell217$ field.\label{l217}}
\end{figure*}

\section{List of CO clumps}\label{clumplist}
\begin{deluxetable}{lccccccccc}
\tabletypesize{\scriptsize}
\tablecaption{List of the clumps detected in the CO(1-0) line\label{COclumps}}
\tablewidth{0pt}
\tablehead{
\colhead{ID} & \colhead{$\ell$} & \colhead{$b$} & \colhead{$v_{lsr}$} & \colhead{$\sigma_v$} & 
\colhead{$d$} & \colhead{$r$} & \colhead{$M$} & \colhead{$M_{vir}$} & \colhead{$L$} \\
 & \colhead{$[^{\circ}]$} & \colhead{$[^{\circ}]$} & \colhead{[km s$^{-1}]$} & \colhead{[km s$^{-1}$]} & \colhead{[kpc]} & \colhead{[pc]} & 
\colhead{$[M_{\odot}]$} & \colhead{$[M_{\odot}]$} & \colhead{[K~km~s$^{-1}$~pc$^2$]} }
\startdata
       1 &   216.61829 &  -1.82868 &  44.0 &   1.8 &   4.3 &   5.7 &  5770. & 19154. &   704.6 \\
       2 &   216.62572 &  -1.64908 &  30.6 &   0.9 &   2.6 &   6.5 &  2707. &  5009. &   366.3 \\
       3 &   216.64683 &  -2.24570 &  23.8 &   1.9 &   1.9 &   3.6 &  3943. & 13551. &   554.5 \\
       4 &   216.66404 &  -2.08390 &  25.9 &   2.0 &   2.1 &   5.0 &  6811. & 20697. &   947.4 \\
       5 &   216.66706 &  -1.70928 &  32.8 &   0.8 &   2.8 &   8.9 &  5936. &  6110. &   791.9 \\
       6 &   216.67371 &  -1.42988 &  30.9 &   1.1 &   2.6 &   5.5 &  3611. &  7170. &   487.8 \\
       7 &   216.68618 &  -0.48407 &  55.2 &   1.7 &   6.1 &   8.7 &  6204. & 25071. &   671.9 \\
       8 &   216.69400 &  -1.88198 &  29.7 &   1.6 &   2.5 &   7.8 &  8458. & 19509. &  1151.3 \\
\enddata
\end{deluxetable}

\section{List of Herschel sources}\label{higallist}
\begin{deluxetable}{lcccccccccccccc}
\rotate
\tabletypesize{\scriptsize}
\tablecaption{List of the analyzed Hi-GAL sources\label{higaltab}}
\tablewidth{0pt}
\tablehead{
\colhead{ID} & \colhead{$\ell$} & \colhead{$b$} & \colhead{$d$} & \colhead{$F_{70}\pm\Delta F_{70}$} & 
\colhead{$F_{160}\pm\Delta F_{160}$} & \colhead{$F_{250}\pm\Delta F_{250}$} & \colhead{$F_{350}\pm\Delta F_{350}$} & 
\colhead{$F_{500}\pm\Delta F_{500}$} & \colhead{$r$} & \colhead{$M$} & \colhead{$T$} & \colhead{$F_{22}$} & \colhead{$L_{bol}$} & unbound/pre-/proto-stellar\\
 & \colhead{$[^{\circ}]$} & \colhead{$[^{\circ}]$} & \colhead{[kpc]} & \colhead{[Jy]} & \colhead{[Jy]} & \colhead{[Jy]} & \colhead{[Jy]} & \colhead{[Jy]} & 
\colhead{$[{\arcsec}]$} & \colhead{[$M_{\odot}$]} & \colhead{[K]} & \colhead{[Jy]} & \colhead{[$L_{\odot}$]} & 0/1/2}
\startdata
    1 &   216.66360 &    -1.66179 & 2.6 &              \nodata &              \nodata &      0.43$\pm$  0.04 &      0.28$\pm$  0.02 &      0.17$\pm$  0.03 &    21.0 &      6.4 &  11.9 &    \nodata &       0.4 &  1 \\
    2 &   216.68802 &    -1.42173 & 2.6 &              \nodata &              \nodata &      0.29$\pm$  0.07 &      0.29$\pm$  0.01 &      0.12$\pm$  0.02 &    13.6 &      5.9 &  11.5 &    \nodata &       0.3 &  1 \\
    3 &   216.69028 &    -1.37625 & 2.6 &              \nodata &              \nodata &      0.31$\pm$  0.09 &      0.27$\pm$  0.02 &      0.23$\pm$  0.01 &    21.9 &     17.5 &   9.3 &    \nodata &       0.4 &  1 \\
    4 &   216.69757 &    -1.41449 & 2.6 &              \nodata &              \nodata &      0.89$\pm$  0.07 &      0.79$\pm$  0.01 &      0.51$\pm$  0.05 &    31.0 &     33.4 &  10.1 &    \nodata &       1.1 &  1 \\
    5 &   216.74443 &    -0.39604 & 2.1 &              \nodata &              \nodata &      0.46$\pm$  0.11 &      0.52$\pm$  0.01 &      0.27$\pm$  0.04 &    25.6 &     13.4 &   9.9 &    \nodata &       0.4 &  1 \\
    6 &   216.78496 &    -1.06081 & 2.6 &      0.87$\pm$  0.03 &      2.45$\pm$  0.12 &      3.99$\pm$  0.15 &      2.61$\pm$  0.02 &      1.36$\pm$  0.01 &    16.2 &     53.1 &  12.1 &       0.01 &      17.2 &  2 \\
    7 &   216.78586 &    -1.04352 & 2.6 &     24.01$\pm$  0.14 &     26.32$\pm$  0.24 &     20.11$\pm$  0.18 &      8.66$\pm$  0.02 &      3.61$\pm$  0.01 &    19.4 &     55.6 &  18.1 &       3.18 &     322.1 &  2 \\
    8 &   216.79010 &    -1.33692 & 2.7 &              \nodata &              \nodata &      0.35$\pm$  0.06 &      0.30$\pm$  0.01 &      0.19$\pm$  0.01 &    16.7 &     12.2 &  10.4 &    \nodata &       0.4 &  1 \\
\enddata
\end{deluxetable}


\begin{thebibliography}{96}
\expandafter\ifx\csname natexlab\endcsname\relax\def\natexlab#1{#1}\fi

\bibitem[{{Alves} {et~al.}(2007){Alves}, {Lombardi}, \& {Lada}}]{alv07}
{Alves}, J., {Lombardi}, M., \& {Lada}, C.~J. 2007, \aap, 462, L17

\bibitem[{{Andr{\'e}} {et~al.}(2000){Andr{\'e}}, {Ward-Thompson}, \&
  {Barsony}}]{and00}
{Andr{\'e}}, P., {Ward-Thompson}, D., \& {Barsony}, M. 2000, Protostars and
  Planets IV, 59

\bibitem[{{Andr{\'e}} {et~al.}(2010){Andr{\'e}}, {Men'shchikov}, {Bontemps},
  {K{\"o}nyves}, {Motte}, {Schneider}, {Didelon}, {Minier}, {Saraceno},
  {Ward-Thompson}, {di Francesco}, {White}, {Molinari}, {Testi}, {Abergel},
  {Griffin}, {Henning}, {Royer}, {Mer{\'{\i}}n}, {Vavrek}, {Attard},
  {Arzoumanian}, {Wilson}, {Ade}, {Aussel}, {Baluteau}, {Benedettini},
  {Bernard}, {Blommaert}, {Cambr{\'e}sy}, {Cox}, {di Giorgio}, {Hargrave},
  {Hennemann}, {Huang}, {Kirk}, {Krause}, {Launhardt}, {Leeks}, {Le Pennec},
  {Li}, {Martin}, {Maury}, {Olofsson}, {Omont}, {Peretto}, {Pezzuto}, {Prusti},
  {Roussel}, {Russeil}, {Sauvage}, {Sibthorpe}, {Sicilia-Aguilar}, {Spinoglio},
  {Waelkens}, {Woodcraft}, \& {Zavagno}}]{and10}
{Andr{\'e}}, P., {Men'shchikov}, A., {Bontemps}, S., {et~al.} 2010, \aap, 518,
  L102

\bibitem[{{Bergin} \& {Tafalla}(2007)}]{ber07}
{Bergin}, E.~A., \& {Tafalla}, M. 2007, \araa, 45, 339

\bibitem[{{Bernard} {et~al.}(2010){Bernard}, {Paradis}, {Marshall}, {Montier},
  {Lagache}, {Paladini}, {Veneziani}, {Brunt}, {Mottram}, {Martin},
  {Ristorcelli}, {Noriega-Crespo}, {Compi{\`e}gne}, {Flagey}, {Anderson},
  {Popescu}, {Tuffs}, {Reach}, {White}, {Benedetti}, {Calzoletti}, {Digiorgio},
  {Faustini}, {Juvela}, {Joblin}, {Joncas}, {Mivilles-Deschenes}, {Olmi},
  {Traficante}, {Piacentini}, {Zavagno}, \& {Molinari}}]{ber10}
{Bernard}, J.-P., {Paradis}, D., {Marshall}, D.~J., {et~al.} 2010, \aap, 518,
  L88

\bibitem[{{Blitz} {et~al.}(1982){Blitz}, {Fich}, \& {Stark}}]{bli82}
{Blitz}, L., {Fich}, M., \& {Stark}, A.~A. 1982, \apjs, 49, 183

\bibitem[{{Bolatto} {et~al.}(2008){Bolatto}, {Leroy}, {Rosolowsky}, {Walter},
  \& {Blitz}}]{bol08}
{Bolatto}, A.~D., {Leroy}, A.~K., {Rosolowsky}, E., {Walter}, F., \& {Blitz},
  L. 2008, \apj, 686, 948

\bibitem[{{Bontemps} {et~al.}(2010){Bontemps}, {Andr{\'e}}, {K{\"o}nyves},
  {Men'shchikov}, {Schneider}, {Maury}, {Peretto}, {Arzoumanian}, {Attard},
  {Motte}, {Minier}, {Didelon}, {Saraceno}, {Abergel}, {Baluteau}, {Bernard},
  {Cambr{\'e}sy}, {Cox}, {di Francesco}, {di Giorgo}, {Griffin}, {Hargrave},
  {Huang}, {Kirk}, {Li}, {Martin}, {Mer{\'{\i}}n}, {Molinari}, {Olofsson},
  {Pezzuto}, {Prusti}, {Roussel}, {Russeil}, {Sauvage}, {Sibthorpe},
  {Spinoglio}, {Testi}, {Vavrek}, {Ward-Thompson}, {White}, {Wilson},
  {Woodcraft}, \& {Zavagno}}]{bon10}
{Bontemps}, S., {Andr{\'e}}, P., {K{\"o}nyves}, V., {et~al.} 2010, \aap, 518,
  L85

\bibitem[{{Brand} {et~al.}(2001){Brand}, {Cesaroni}, {Palla}, \&
  {Molinari}}]{bra01}
{Brand}, J., {Cesaroni}, R., {Palla}, F., \& {Molinari}, S. 2001, \aap, 370,
  230

\bibitem[{{Brunt} {et~al.}(2003){Brunt}, {Kerton}, \& {Pomerleau}}]{bru03}
{Brunt}, C.~M., {Kerton}, C.~R., \& {Pomerleau}, C. 2003, \apjs, 144, 47

\bibitem[{{Brunthaler} {et~al.}(2011){Brunthaler}, {Reid}, {Menten}, {Zheng},
  {Bartkiewicz}, {Choi}, {Dame}, {Hachisuka}, {Immer}, {Moellenbrock},
  {Moscadelli}, {Rygl}, {Sanna}, {Sato}, {Wu}, {Xu}, \& {Zhang}}]{bru11}
{Brunthaler}, A., {Reid}, M.~J., {Menten}, K.~M., {et~al.} 2011, Astronomische
  Nachrichten, 332, 461

\bibitem[{{Burton} \& {te Lintel Hekkert}(1986)}]{bur86}
{Burton}, W.~B., \& {te Lintel Hekkert}, P. 1986, \aaps, 65, 427

\bibitem[{{Carlhoff} {et~al.}(submitted){Carlhoff}, {Nguyen Luong}, {Schilke},
  \& {et al.}}]{car13}
{Carlhoff}, P., {Nguyen Luong}, Q., {Schilke}, P., \& {et al.} submitted, \aap

\bibitem[{{Carraro} {et~al.}(2005){Carraro}, {V{\'a}zquez}, {Moitinho}, \&
  {Baume}}]{car05}
{Carraro}, G., {V{\'a}zquez}, R.~A., {Moitinho}, A., \& {Baume}, G. 2005,
  \apjl, 630, L153

\bibitem[{{Chabrier}(2005)}]{cha05}
{Chabrier}, G. 2005, in Astrophysics and Space Science Library, Vol. 327, The
  Initial Mass Function 50 Years Later, ed. E.~{Corbelli}, F.~{Palla}, \&
  H.~{Zinnecker}, 41

\bibitem[{{Clari{\'a}}(1974)}]{cla74}
{Clari{\'a}}, J.~J. 1974, \aap, 37, 229

\bibitem[{{Cordes} \& {Lazio}(2002)}]{cor02}
{Cordes}, J.~M., \& {Lazio}, T.~J.~W. 2002, ArXiv Astrophysics e-prints

\bibitem[{{Dame} {et~al.}(2001){Dame}, {Hartmann}, \& {Thaddeus}}]{dam01}
{Dame}, T.~M., {Hartmann}, D., \& {Thaddeus}, P. 2001, \apj, 547, 792

\bibitem[{{Dobashi} {et~al.}(2005){Dobashi}, {Uehara}, {Kandori}, {Sakurai},
  {Kaiden}, {Umemoto}, \& {Sato}}]{dob05}
{Dobashi}, K., {Uehara}, H., {Kandori}, R., {et~al.} 2005, \pasj, 57, 1

\bibitem[{{Egan} {et~al.}(1998){Egan}, {Shipman}, {Price}, {Carey}, {Clark}, \&
  {Cohen}}]{ega98}
{Egan}, M.~P., {Shipman}, R.~F., {Price}, S.~D., {et~al.} 1998, \apjl, 494,
  L199

\bibitem[{{Elia} {et~al.}(2007){Elia}, {Massi}, {Strafella}, {De Luca},
  {Giannini}, {Lorenzetti}, {Nisini}, {Campeggio}, \& {Maiolo}}]{eli07}
{Elia}, D., {Massi}, F., {Strafella}, F., {et~al.} 2007, \apj, 655, 316

\bibitem[{{Elia} {et~al.}(2010){Elia}, {Schisano}, {Molinari}, {Robitaille},
  {Angl{\'e}s-Alc{\'a}zar}, {Bally}, {Battersby}, {Benedettini}, {Billot},
  {Calzoletti}, {di Giorgio}, {Faustini}, {Li}, {Martin}, {Morgan}, {Motte},
  {Mottram}, {Natoli}, {Olmi}, {Paladini}, {Piacentini}, {Pestalozzi},
  {Pezzuto}, {Polychroni}, {Smith}, {Strafella}, {Stringfellow}, {Testi},
  {Thompson}, {Traficante}, \& {Veneziani}}]{eli10}
{Elia}, D., {Schisano}, E., {Molinari}, S., {et~al.} 2010, \aap, 518, L97

\bibitem[{{Enoch} {et~al.}(2008){Enoch}, {Evans}, {Sargent}, {Glenn},
  {Rosolowsky}, \& {Myers}}]{eno08}
{Enoch}, M.~L., {Evans}, II, N.~J., {Sargent}, A.~I., {et~al.} 2008, \apj, 684,
  1240

\bibitem[{{Federrath} {et~al.}(2008){Federrath}, {Glover}, {Klessen}, \&
  {Schmidt}}]{fed08}
{Federrath}, C., {Glover}, S.~C.~O., {Klessen}, R.~S., \& {Schmidt}, W. 2008,
  Physica Scripta Volume T, 132, 014025

\bibitem[{{Giannini} {et~al.}(2012){Giannini}, {Elia}, {Lorenzetti},
  {Molinari}, {Motte}, {Schisano}, {Pezzuto}, {Pestalozzi}, {di Giorgio},
  {Andr{\'e}}, {Hill}, {Benedettini}, {Bontemps}, {di Francesco}, {Fallscheer},
  {Hennemann}, {Kirk}, {Minier}, {Nguyen Luong}, {Polychroni}, {Rygl},
  {Saraceno}, {Schneider}, {Spinoglio}, {Testi}, {Ward-Thompson}, \&
  {White}}]{gia12}
{Giannini}, T., {Elia}, D., {Lorenzetti}, D., {et~al.} 2012, \aap, 539, A156

\bibitem[{{Gregorio-Hetem} {et~al.}(2009){Gregorio-Hetem}, {Montmerle},
  {Rodrigues}, {Marciotto}, {Preibisch}, \& {Zinnecker}}]{gre09}
{Gregorio-Hetem}, J., {Montmerle}, T., {Rodrigues}, C.~V., {et~al.} 2009, \aap,
  506, 711

\bibitem[{{Griffin} {et~al.}(2010){Griffin}, {Abergel}, {Abreu}, {Ade},
  {Andr{\'e}}, {Augueres}, {Babbedge}, {Bae}, {Baillie}, {Baluteau}, {Barlow},
  {Bendo}, {Benielli}, {Bock}, {Bonhomme}, {Brisbin}, {Brockley-Blatt},
  {Caldwell}, {Cara}, {Castro-Rodriguez}, {Cerulli}, {Chanial}, {Chen},
  {Clark}, {Clements}, {Clerc}, {Coker}, {Communal}, {Conversi}, {Cox},
  {Crumb}, {Cunningham}, {Daly}, {Davis}, {de Antoni}, {Delderfield}, {Devin},
  {di Giorgio}, {Didschuns}, {Dohlen}, {Donati}, {Dowell}, {Dowell}, {Duband},
  {Dumaye}, {Emery}, {Ferlet}, {Ferrand}, {Fontignie}, {Fox}, {Franceschini},
  {Frerking}, {Fulton}, {Garcia}, {Gastaud}, {Gear}, {Glenn}, {Goizel},
  {Griffin}, {Grundy}, {Guest}, {Guillemet}, {Hargrave}, {Harwit}, {Hastings},
  {Hatziminaoglou}, {Herman}, {Hinde}, {Hristov}, {Huang}, {Imhof}, {Isaak},
  {Israelsson}, {Ivison}, {Jennings}, {Kiernan}, {King}, {Lange}, {Latter},
  {Laurent}, {Laurent}, {Leeks}, {Lellouch}, {Levenson}, {Li}, {Li},
  {Lilienthal}, {Lim}, {Liu}, {Lu}, {Madden}, {Mainetti}, {Marliani}, {McKay},
  {Mercier}, {Molinari}, {Morris}, {Moseley}, {Mulder}, {Mur}, {Naylor},
  {Nguyen}, {O'Halloran}, {Oliver}, {Olofsson}, {Olofsson}, {Orfei}, {Page},
  {Pain}, {Panuzzo}, {Papageorgiou}, {Parks}, {Parr-Burman}, {Pearce},
  {Pearson}, {P{\'e}rez-Fournon}, {Pinsard}, {Pisano}, {Podosek}, {Pohlen},
  {Polehampton}, {Pouliquen}, {Rigopoulou}, {Rizzo}, {Roseboom}, {Roussel},
  {Rowan-Robinson}, {Rownd}, {Saraceno}, {Sauvage}, {Savage}, {Savini},
  {Sawyer}, {Scharmberg}, {Schmitt}, {Schneider}, {Schulz}, {Schwartz},
  {Shafer}, {Shupe}, {Sibthorpe}, {Sidher}, {Smith}, {Smith}, {Smith},
  {Spencer}, {Stobie}, {Sudiwala}, {Sukhatme}, {Surace}, {Stevens}, {Swinyard},
  {Trichas}, {Tourette}, {Triou}, {Tseng}, {Tucker}, {Turner}, {Vaccari},
  {Valtchanov}, {Vigroux}, {Virique}, {Voellmer}, {Walker}, {Ward}, {Waskett},
  {Weilert}, {Wesson}, {White}, {Whitehouse}, {Wilson}, {Winter}, {Woodcraft},
  {Wright}, {Xu}, {Zavagno}, {Zemcov}, {Zhang}, \& {Zonca}}]{gri10}
{Griffin}, M.~J., {Abergel}, A., {Abreu}, A., {et~al.} 2010, \aap, 518, L3

\bibitem[{{Hennemann} {et~al.}(2010){Hennemann}, {Motte}, {Bontemps},
  {Schneider}, {Csengeri}, {Balog}, {di Francesco}, {Zavagno}, {Andr{\'e}},
  {Men'shchikov}, {Abergel}, {Ali}, {Baluteau}, {Bernard}, {Cox}, {Didelon},
  {di Giorgio}, {Griffin}, {Hargrave}, {Hill}, {Horeau}, {Huang}, {Kirk},
  {Leeks}, {Li}, {Marston}, {Martin}, {Molinari}, {Nguyen Luong}, {Olofsson},
  {Persi}, {Pezzuto}, {Russeil}, {Saraceno}, {Sauvage}, {Sibthorpe},
  {Spinoglio}, {Testi}, {Ward-Thompson}, {White}, {Wilson}, \&
  {Woodcraft}}]{hen10}
{Hennemann}, M., {Motte}, F., {Bontemps}, S., {et~al.} 2010, \aap, 518, L84

\bibitem[{{Heyer} {et~al.}(1998){Heyer}, {Brunt}, {Snell}, {Howe}, {Schloerb},
  \& {Carpenter}}]{hey98}
{Heyer}, M.~H., {Brunt}, C., {Snell}, R.~L., {et~al.} 1998, \apjs, 115, 241

\bibitem[{{Heyer} {et~al.}(2006){Heyer}, {Williams}, \& {Brunt}}]{hey06}
{Heyer}, M.~H., {Williams}, J.~P., \& {Brunt}, C.~M. 2006, \apj, 643, 956

\bibitem[{{Hildebrand}(1983)}]{hil83}
{Hildebrand}, R.~H. 1983, \qjras, 24, 267

\bibitem[{{Hill} {et~al.}(2011){Hill}, {Motte}, {Didelon}, {Bontemps},
  {Minier}, {Hennemann}, {Schneider}, {Andr{\'e}}, {Men'shchikov}, {Anderson},
  {Arzoumanian}, {Bernard}, {di Francesco}, {Elia}, {Giannini}, {Griffin},
  {K{\"o}nyves}, {Kirk}, {Marston}, {Martin}, {Molinari}, {Nguyen Luong},
  {Peretto}, {Pezzuto}, {Roussel}, {Sauvage}, {Sousbie}, {Testi},
  {Ward-Thompson}, {White}, {Wilson}, \& {Zavagno}}]{hil11}
{Hill}, T., {Motte}, F., {Didelon}, P., {et~al.} 2011, \aap, 533, A94

\bibitem[{{Ikeda} {et~al.}(2007){Ikeda}, {Sunada}, \& {Kitamura}}]{ike07}
{Ikeda}, N., {Sunada}, K., \& {Kitamura}, Y. 2007, \apj, 665, 1194

\bibitem[{{Kainulainen} {et~al.}(2011){Kainulainen}, {Beuther}, {Banerjee},
  {Federrath}, \& {Henning}}]{kai11}
{Kainulainen}, J., {Beuther}, H., {Banerjee}, R., {Federrath}, C., \&
  {Henning}, T. 2011, \aap, 530, A64

\bibitem[{{Kauffmann} \& {Pillai}(2010)}]{kau10}
{Kauffmann}, J., \& {Pillai}, T. 2010, \apjl, 723, L7

\bibitem[{{Kim} {et~al.}(2004){Kim}, {Kawamura}, {Yonekura}, \&
  {Fukui}}]{kim04}
{Kim}, B.~G., {Kawamura}, A., {Yonekura}, Y., \& {Fukui}, Y. 2004, \pasj, 56,
  313

\bibitem[{{Klessen}(2000)}]{kle00}
{Klessen}, R.~S. 2000, \apj, 535, 869

\bibitem[{{K{\"o}nyves} {et~al.}(2010){K{\"o}nyves}, {Andr{\'e}},
  {Men'shchikov}, {Schneider}, {Arzoumanian}, {Bontemps}, {Attard}, {Motte},
  {Didelon}, {Maury}, {Abergel}, {Ali}, {Baluteau}, {Bernard}, {Cambr{\'e}sy},
  {Cox}, {di Francesco}, {di Giorgio}, {Griffin}, {Hargrave}, {Huang}, {Kirk},
  {Li}, {Martin}, {Minier}, {Molinari}, {Olofsson}, {Pezzuto}, {Russeil},
  {Roussel}, {Saraceno}, {Sauvage}, {Sibthorpe}, {Spinoglio}, {Testi},
  {Ward-Thompson}, {White}, {Wilson}, {Woodcraft}, \& {Zavagno}}]{kon10}
{K{\"o}nyves}, V., {Andr{\'e}}, P., {Men'shchikov}, A., {et~al.} 2010, \aap,
  518, L106

\bibitem[{{Kramer} {et~al.}(1998){Kramer}, {Stutzki}, {Rohrig}, \&
  {Corneliussen}}]{kra98}
{Kramer}, C., {Stutzki}, J., {Rohrig}, R., \& {Corneliussen}, U. 1998, \aap,
  329, 249

\bibitem[{{Kroupa}(2001)}]{kro01}
{Kroupa}, P. 2001, \mnras, 322, 231

\bibitem[{{Krumholz} \& {McKee}(2008)}]{kru08}
{Krumholz}, M.~R., \& {McKee}, C.~F. 2008, \nat, 451, 1082

\bibitem[{{Kutner} \& {Ulich}(1981)}]{kut81}
{Kutner}, M.~L., \& {Ulich}, B.~L. 1981, \apj, 250, 341

\bibitem[{{Lada} {et~al.}(2010){Lada}, {Lombardi}, \& {Alves}}]{lad10}
{Lada}, C.~J., {Lombardi}, M., \& {Alves}, J.~F. 2010, \apj, 724, 687

\bibitem[{{Larson}(1981)}]{lar81}
{Larson}, R.~B. 1981, \mnras, 194, 809

\bibitem[{{Lee} {et~al.}(1994){Lee}, {Snell}, \& {Dickman}}]{lee94}
{Lee}, Y., {Snell}, R.~L., \& {Dickman}, R.~L. 1994, \apj, 432, 167

\bibitem[{{Lynds}(1962)}]{lyn62}
{Lynds}, B.~T. 1962, \apjs, 7, 1

\bibitem[{{Maddalena} {et~al.}(1986){Maddalena}, {Morris}, {Moscowitz}, \&
  {Thaddeus}}]{mad86}
{Maddalena}, R.~J., {Morris}, M., {Moscowitz}, J., \& {Thaddeus}, P. 1986,
  \apj, 303, 375

\bibitem[{{May} {et~al.}(1997){May}, {Alvarez}, \& {Bronfman}}]{may97}
{May}, J., {Alvarez}, H., \& {Bronfman}, L. 1997, \aap, 327, 325

\bibitem[{{May} {et~al.}(1993){May}, {Bronfman}, {Alvarez}, {Murphy}, \&
  {Thaddeus}}]{may93}
{May}, J., {Bronfman}, L., {Alvarez}, H., {Murphy}, D.~C., \& {Thaddeus}, P.
  1993, \aaps, 99, 105

\bibitem[{{May} {et~al.}(1988){May}, {Murphy}, \& {Thaddeus}}]{may88}
{May}, J., {Murphy}, D.~C., \& {Thaddeus}, P. 1988, \aaps, 73, 51

\bibitem[{{Megeath} {et~al.}(2009){Megeath}, {Allgaier}, {Young}, {Allen},
  {Pipher}, \& {Wilson}}]{meg09}
{Megeath}, S.~T., {Allgaier}, E., {Young}, E., {et~al.} 2009, \aj, 137, 4072

\bibitem[{{Miyazaki} \& {Tsuboi}(2000)}]{miy00}
{Miyazaki}, A., \& {Tsuboi}, M. 2000, \apj, 536, 357

\bibitem[{{Mizuno} \& {Fukui}(2004)}]{miz04}
{Mizuno}, A., \& {Fukui}, Y. 2004, in Astronomical Society of the Pacific
  Conference Series, Vol. 317, Milky Way Surveys: The Structure and Evolution
  of our Galaxy, ed. D.~{Clemens}, R.~{Shah}, \& T.~{Brainerd}, 59

\bibitem[{{Moitinho} {et~al.}(2006){Moitinho}, {V{\'a}zquez}, {Carraro},
  {Baume}, {Giorgi}, \& {Lyra}}]{moi06}
{Moitinho}, A., {V{\'a}zquez}, R.~A., {Carraro}, G., {et~al.} 2006, \mnras,
  368, L77

\bibitem[{{Molinari} {et~al.}(2008){Molinari}, {Pezzuto}, {Cesaroni}, {Brand},
  {Faustini}, \& {Testi}}]{mol08}
{Molinari}, S., {Pezzuto}, S., {Cesaroni}, R., {et~al.} 2008, \aap, 481, 345

\bibitem[{{Molinari} {et~al.}(2011){Molinari}, {Schisano}, {Faustini},
  {Pestalozzi}, {di Giorgio}, \& {Liu}}]{mol11}
{Molinari}, S., {Schisano}, E., {Faustini}, F., {et~al.} 2011, \aap, 530, A133

\bibitem[{{Molinari} {et~al.}(2010{\natexlab{a}}){Molinari}, {Swinyard},
  {Bally}, {Barlow}, {Bernard}, {Martin}, {Moore}, {Noriega-Crespo}, {Plume},
  {Testi}, {Zavagno}, {Abergel}, {Ali}, {Anderson}, {Andr{\'e}}, {Baluteau},
  {Battersby}, {Beltr{\'a}n}, {Benedettini}, {Billot}, {Blommaert}, {Bontemps},
  {Boulanger}, {Brand}, {Brunt}, {Burton}, {Calzoletti}, {Carey}, {Caselli},
  {Cesaroni}, {Cernicharo}, {Chakrabarti}, {Chrysostomou}, {Cohen},
  {Compiegne}, {de Bernardis}, {de Gasperis}, {di Giorgio}, {Elia}, {Faustini},
  {Flagey}, {Fukui}, {Fuller}, {Ganga}, {Garcia-Lario}, {Glenn}, {Goldsmith},
  {Griffin}, {Hoare}, {Huang}, {Ikhenaode}, {Joblin}, {Joncas}, {Juvela},
  {Kirk}, {Lagache}, {Li}, {Lim}, {Lord}, {Marengo}, {Marshall}, {Masi},
  {Massi}, {Matsuura}, {Minier}, {Miville-Desch{\^e}nes}, {Montier}, {Morgan},
  {Motte}, {Mottram}, {M{\"u}ller}, {Natoli}, {Neves}, {Olmi}, {Paladini},
  {Paradis}, {Parsons}, {Peretto}, {Pestalozzi}, {Pezzuto}, {Piacentini},
  {Piazzo}, {Polychroni}, {Pomar{\`e}s}, {Popescu}, {Reach}, {Ristorcelli},
  {Robitaille}, {Robitaille}, {Rod{\'o}n}, {Roy}, {Royer}, {Russeil},
  {Saraceno}, {Sauvage}, {Schilke}, {Schisano}, {Schneider}, {Schuller},
  {Schulz}, {Sibthorpe}, {Smith}, {Smith}, {Spinoglio}, {Stamatellos},
  {Strafella}, {Stringfellow}, {Sturm}, {Taylor}, {Thompson}, {Traficante},
  {Tuffs}, {Umana}, {Valenziano}, {Vavrek}, {Veneziani}, {Viti}, {Waelkens},
  {Ward-Thompson}, {White}, {Wilcock}, {Wyrowski}, {Yorke}, \&
  {Zhang}}]{mol10b}
{Molinari}, S., {Swinyard}, B., {Bally}, J., {et~al.} 2010{\natexlab{a}}, \aap,
  518, L100

\bibitem[{{Molinari} {et~al.}(2010{\natexlab{b}}){Molinari}, {Swinyard},
  {Bally}, {Barlow}, {Bernard}, {Martin}, {Moore}, {Noriega-Crespo}, {Plume},
  {Testi}, {Zavagno}, {Abergel}, {Ali}, {Andr{\'e}}, {Baluteau}, {Benedettini},
  {Bern{\'e}}, {Billot}, {Blommaert}, {Bontemps}, {Boulanger}, {Brand},
  {Brunt}, {Burton}, {Campeggio}, {Carey}, {Caselli}, {Cesaroni}, {Cernicharo},
  {Chakrabarti}, {Chrysostomou}, {Codella}, {Cohen}, {Compiegne}, {Davis}, {de
  Bernardis}, {de Gasperis}, {Di Francesco}, {di Giorgio}, {Elia}, {Faustini},
  {Fischera}, {Fukui}, {Fuller}, {Ganga}, {Garcia-Lario}, {Giard}, {Giardino},
  {Glenn}, {Goldsmith}, {Griffin}, {Hoare}, {Huang}, {Jiang}, {Joblin},
  {Joncas}, {Juvela}, {Kirk}, {Lagache}, {Li}, {Lim}, {Lord}, {Lucas},
  {Maiolo}, {Marengo}, {Marshall}, {Masi}, {Massi}, {Matsuura}, {Meny},
  {Minier}, {Miville-Desch{\^e}nes}, {Montier}, {Motte}, {M{\"u}ller},
  {Natoli}, {Neves}, {Olmi}, {Paladini}, {Paradis}, {Pestalozzi}, {Pezzuto},
  {Piacentini}, {Pomar{\`e}s}, {Popescu}, {Reach}, {Richer}, {Ristorcelli},
  {Roy}, {Royer}, {Russeil}, {Saraceno}, {Sauvage}, {Schilke},
  {Schneider-Bontemps}, {Schuller}, {Schultz}, {Shepherd}, {Sibthorpe},
  {Smith}, {Smith}, {Spinoglio}, {Stamatellos}, {Strafella}, {Stringfellow},
  {Sturm}, {Taylor}, {Thompson}, {Tuffs}, {Umana}, {Valenziano}, {Vavrek},
  {Viti}, {Waelkens}, {Ward-Thompson}, {White}, {Wyrowski}, {Yorke}, \&
  {Zhang}}]{mol10a}
---. 2010{\natexlab{b}}, \pasp, 122, 314

\bibitem[{{Motte} \& {Andr{\'e}}(2001)}]{mot01}
{Motte}, F., \& {Andr{\'e}}, P. 2001, \aap, 365, 440

\bibitem[{{Motte} {et~al.}(2010){Motte}, {Zavagno}, {Bontemps}, {Schneider},
  {Hennemann}, {di Francesco}, {Andr{\'e}}, {Saraceno}, {Griffin}, {Marston},
  {Ward-Thompson}, {White}, {Minier}, {Men'shchikov}, {Hill}, {Abergel},
  {Anderson}, {Aussel}, {Balog}, {Baluteau}, {Bernard}, {Cox}, {Csengeri},
  {Deharveng}, {Didelon}, {di Giorgio}, {Hargrave}, {Huang}, {Kirk}, {Leeks},
  {Li}, {Martin}, {Molinari}, {Nguyen-Luong}, {Olofsson}, {Persi}, {Peretto},
  {Pezzuto}, {Roussel}, {Russeil}, {Sadavoy}, {Sauvage}, {Sibthorpe},
  {Spinoglio}, {Testi}, {Teyssier}, {Vavrek}, {Wilson}, \& {Woodcraft}}]{mot10}
{Motte}, F., {Zavagno}, A., {Bontemps}, S., {et~al.} 2010, \aap, 518, L77

\bibitem[{{Murray}(2011)}]{mur11}
{Murray}, N. 2011, \apj, 729, 133

\bibitem[{{Nakanishi} \& {Sofue}(2006)}]{nak06}
{Nakanishi}, H., \& {Sofue}, Y. 2006, \pasj, 58, 847

\bibitem[{{Ogawa} {et~al.}(1990){Ogawa}, {Mizuno}, {Ishikawa}, {Fukui}, \&
  {Hoko}}]{oga90}
{Ogawa}, H., {Mizuno}, A., {Ishikawa}, H., {Fukui}, Y., \& {Hoko}, H. 1990,
  International Journal of Infrared and Millimeter Waves, 11, 717

\bibitem[{{Olmi} {et~al.}(2010){Olmi}, {Angl{\'e}s-Alc{\'a}zar}, {De Luca},
  {Elia}, {Giannini}, {Lorenzetti}, {Massi}, {Martin}, \& {Strafella}}]{olm10}
{Olmi}, L., {Angl{\'e}s-Alc{\'a}zar}, D., {De Luca}, M., {et~al.} 2010, \apj,
  723, 1065

\bibitem[{{Olmi} {et~al.}(2013){Olmi}, {Angl{\'e}s-Alc{\'a}zar}, {Elia},
  {Molinari}, {Montier}, {Pestalozzi}, {Pezzuto}, {Polychroni}, {Ristorcelli},
  {Rodon}, {Schisano}, {Smith}, {Testi}, \& {Thompson}}]{olm13}
{Olmi}, L., {Angl{\'e}s-Alc{\'a}zar}, D., {Elia}, D., {et~al.} 2013, \aap, 551,
  A111

\bibitem[{{Ott}(2010)}]{ott10}
{Ott}, S. 2010, in Astronomical Society of the Pacific Conference Series, Vol.
  434, Astronomical Data Analysis Software and Systems XIX, ed. Y.~{Mizumoto},
  K.-I. {Morita}, \& M.~{Ohishi}, 139

\bibitem[{Piazzo {et~al.}(2012)Piazzo, Ikhenaode, Natoli, Pestalozzi,
  Piacentini, \& Traficante}]{pia12}
Piazzo, L., Ikhenaode, D., Natoli, P., {et~al.} 2012, Image Processing, IEEE
  Transactions on, 21, 3687

\bibitem[{{Pilbratt} {et~al.}(2010){Pilbratt}, {Riedinger}, {Passvogel},
  {Crone}, {Doyle}, {Gageur}, {Heras}, {Jewell}, {Metcalfe}, {Ott}, \&
  {Schmidt}}]{pil10}
{Pilbratt}, G.~L., {Riedinger}, J.~R., {Passvogel}, T., {et~al.} 2010, \aap,
  518, L1

\bibitem[{{Pineda} {et~al.}(2008){Pineda}, {Caselli}, \& {Goodman}}]{pin08}
{Pineda}, J.~E., {Caselli}, P., \& {Goodman}, A.~A. 2008, \apj, 679, 481

\bibitem[{{Poglitsch} {et~al.}(2010){Poglitsch}, {Waelkens}, {Geis},
  {Feuchtgruber}, {Vandenbussche}, {Rodriguez}, {Krause}, {Renotte}, {van
  Hoof}, {Saraceno}, {Cepa}, {Kerschbaum}, {Agn{\`e}se}, {Ali}, {Altieri},
  {Andreani}, {Augueres}, {Balog}, {Barl}, {Bauer}, {Belbachir}, {Benedettini},
  {Billot}, {Boulade}, {Bischof}, {Blommaert}, {Callut}, {Cara}, {Cerulli},
  {Cesarsky}, {Contursi}, {Creten}, {De Meester}, {Doublier}, {Doumayrou},
  {Duband}, {Exter}, {Genzel}, {Gillis}, {Gr{\"o}zinger}, {Henning},
  {Herreros}, {Huygen}, {Inguscio}, {Jakob}, {Jamar}, {Jean}, {de Jong},
  {Katterloher}, {Kiss}, {Klaas}, {Lemke}, {Lutz}, {Madden}, {Marquet},
  {Martignac}, {Mazy}, {Merken}, {Montfort}, {Morbidelli}, {M{\"u}ller},
  {Nielbock}, {Okumura}, {Orfei}, {Ottensamer}, {Pezzuto}, {Popesso},
  {Putzeys}, {Regibo}, {Reveret}, {Royer}, {Sauvage}, {Schreiber}, {Stegmaier},
  {Schmitt}, {Schubert}, {Sturm}, {Thiel}, {Tofani}, {Vavrek}, {Wetzstein},
  {Wieprecht}, \& {Wiezorrek}}]{pog10}
{Poglitsch}, A., {Waelkens}, C., {Geis}, N., {et~al.} 2010, \aap, 518, L2

\bibitem[{{Preibisch} {et~al.}(1993){Preibisch}, {Ossenkopf}, {Yorke}, \&
  {Henning}}]{pre93}
{Preibisch}, T., {Ossenkopf}, V., {Yorke}, H.~W., \& {Henning}, T. 1993, \aap,
  279, 577

\bibitem[{{Reid} {et~al.}(2009){Reid}, {Menten}, {Zheng}, {Brunthaler},
  {Moscadelli}, {Xu}, {Zhang}, {Sato}, {Honma}, {Hirota}, {Hachisuka}, {Choi},
  {Moellenbrock}, \& {Bartkiewicz}}]{rei09}
{Reid}, M.~J., {Menten}, K.~M., {Zheng}, X.~W., {et~al.} 2009, \apj, 700, 137

\bibitem[{{Reipurth} \& {Yan}(2008)}]{rei08}
{Reipurth}, B., \& {Yan}, C.-H. 2008, {Star Formation and Molecular Clouds
  towards the Galactic Anti-Center}, ed. B.~{Reipurth}, 869

\bibitem[{{Rosolowsky} \& {Blitz}(2005)}]{ros05}
{Rosolowsky}, E., \& {Blitz}, L. 2005, \apj, 623, 826

\bibitem[{{Rosolowsky} \& {Leroy}(2006)}]{ros06}
{Rosolowsky}, E., \& {Leroy}, A. 2006, \pasp, 118, 590

\bibitem[{{Rosolowsky} \& {Leroy}(2011)}]{ros11}
{Rosolowsky}, E., \& {Leroy}, A. 2011, in Astrophysics Source Code Library,
  record ascl:1102.012, 2012

\bibitem[{{Ruprecht}(1966)}]{rup66}
{Ruprecht}, J. 1966, IAU Trans. XIIB, 348

\bibitem[{{Russeil} {et~al.}(2011){Russeil}, {Pestalozzi}, {Mottram},
  {Bontemps}, {Anderson}, {Zavagno}, {Beltr{\'a}n}, {Bally}, {Brand}, {Brunt},
  {Cesaroni}, {Joncas}, {Marshall}, {Martin}, {Massi}, {Molinari}, {Moore},
  {Noriega-Crespo}, {Olmi}, {Thompson}, {Wienen}, \& {Wyrowski}}]{rus11}
{Russeil}, D., {Pestalozzi}, M., {Mottram}, J.~C., {et~al.} 2011, \aap, 526,
  A151

\bibitem[{{Sadavoy} {et~al.}(2012){Sadavoy}, {di Francesco}, {Andr{\'e}},
  {Pezzuto}, {Bernard}, {Bontemps}, {Bressert}, {Chitsazzadeh}, {Fallscheer},
  {Hennemann}, {Hill}, {Martin}, {Motte}, {Nguyen Luong}, {Peretto}, {Reid},
  {Schneider}, {Testi}, {White}, \& {Wilson}}]{sad12}
{Sadavoy}, S.~I., {di Francesco}, J., {Andr{\'e}}, P., {et~al.} 2012, \aap,
  540, A10

\bibitem[{{Saraceno} {et~al.}(1996){Saraceno}, {Andre}, {Ceccarelli},
  {Griffin}, \& {Molinari}}]{sar96}
{Saraceno}, P., {Andre}, P., {Ceccarelli}, C., {Griffin}, M., \& {Molinari}, S.
  1996, \aap, 309, 827

\bibitem[{{Schneider} {et~al.}(2012){Schneider}, {Csengeri}, {Hennemann},
  {Motte}, {Didelon}, {Federrath}, {Bontemps}, {Di Francesco}, {Arzoumanian},
  {Minier}, {Andr{\'e}}, {Hill}, {Zavagno}, {Nguyen-Luong}, {Attard},
  {Bernard}, {Elia}, {Fallscheer}, {Griffin}, {Kirk}, {Klessen}, {K{\"o}nyves},
  {Martin}, {Men'shchikov}, {Palmeirim}, {Peretto}, {Pestalozzi}, {Russeil},
  {Sadavoy}, {Sousbie}, {Testi}, {Tremblin}, {Ward-Thompson}, \&
  {White}}]{sch12}
{Schneider}, N., {Csengeri}, T., {Hennemann}, M., {et~al.} 2012, \aap, 540, L11

\bibitem[{{Sharpless}(1959)}]{sha59}
{Sharpless}, S. 1959, \apjs, 4, 257

\bibitem[{{Shetty} {et~al.}(2011){Shetty}, {Glover}, {Dullemond}, \&
  {Klessen}}]{she11}
{Shetty}, R., {Glover}, S.~C., {Dullemond}, C.~P., \& {Klessen}, R.~S. 2011,
  \mnras, 412, 1686

\bibitem[{{Simon} {et~al.}(2001){Simon}, {Jackson}, {Clemens}, {Bania}, \&
  {Heyer}}]{sim01}
{Simon}, R., {Jackson}, J.~M., {Clemens}, D.~P., {Bania}, T.~M., \& {Heyer},
  M.~H. 2001, \apj, 551, 747

\bibitem[{{Solomon} {et~al.}(1992){Solomon}, {Downes}, \& {Radford}}]{sol92}
{Solomon}, P.~M., {Downes}, D., \& {Radford}, S.~J.~E. 1992, \apjl, 398, L29

\bibitem[{{Solomon} {et~al.}(1987){Solomon}, {Rivolo}, {Barrett}, \&
  {Yahil}}]{sol87}
{Solomon}, P.~M., {Rivolo}, A.~R., {Barrett}, J., \& {Yahil}, A. 1987, \apj,
  319, 730

\bibitem[{{Tegmark}(1997)}]{teg97}
{Tegmark}, M. 1997, \apjl, 480, L87

\bibitem[{{Traficante} {et~al.}(2011){Traficante}, {Calzoletti}, {Veneziani},
  {Ali}, {de Gasperis}, {di Giorgio}, {Faustini}, {Ikhenaode}, {Molinari},
  {Natoli}, {Pestalozzi}, {Pezzuto}, {Piacentini}, {Piazzo}, {Polenta}, \&
  {Schisano}}]{tra11}
{Traficante}, A., {Calzoletti}, L., {Veneziani}, M., {et~al.} 2011, \mnras,
  416, 2932

\bibitem[{{Vall{\'e}e}(2005)}]{val05}
{Vall{\'e}e}, J.~P. 2005, \aj, 130, 569

\bibitem[{{V{\'a}zquez} {et~al.}(2008){V{\'a}zquez}, {May}, {Carraro},
  {Bronfman}, {Moitinho}, \& {Baume}}]{vaz08}
{V{\'a}zquez}, R.~A., {May}, J., {Carraro}, G., {et~al.} 2008, \apj, 672, 930

\bibitem[{{Veneziani} {et~al.}(2013){Veneziani}, {Elia}, {Noriega-Crespo},
  {Paladini}, {Carey}, {Faimali}, {Molinari}, {Pestalozzi}, {Piacentini},
  {Schisano}, \& {Tibbs}}]{ven13}
{Veneziani}, M., {Elia}, D., {Noriega-Crespo}, A., {et~al.} 2013, \aap, 549,
  A130

\bibitem[{{Williams} {et~al.}(1994){Williams}, {de Geus}, \& {Blitz}}]{wil94}
{Williams}, J.~P., {de Geus}, E.~J., \& {Blitz}, L. 1994, \apj, 428, 693

\bibitem[{{Wouterloot} {et~al.}(1990){Wouterloot}, {Brand}, {Burton}, \&
  {Kwee}}]{wou90}
{Wouterloot}, J.~G.~A., {Brand}, J., {Burton}, W.~B., \& {Kwee}, K.~K. 1990,
  \aap, 230, 21

\bibitem[{{Wright} {et~al.}(2010){Wright}, {Eisenhardt}, {Mainzer}, {Ressler},
  {Cutri}, {Jarrett}, {Kirkpatrick}, {Padgett}, {McMillan}, {Skrutskie},
  {Stanford}, {Cohen}, {Walker}, {Mather}, {Leisawitz}, {Gautier}, {McLean},
  {Benford}, {Lonsdale}, {Blain}, {Mendez}, {Irace}, {Duval}, {Liu}, {Royer},
  {Heinrichsen}, {Howard}, {Shannon}, {Kendall}, {Walsh}, {Larsen}, {Cardon},
  {Schick}, {Schwalm}, {Abid}, {Fabinsky}, {Naes}, \& {Tsai}}]{wri10}
{Wright}, E.~L., {Eisenhardt}, P.~R.~M., {Mainzer}, A.~K., {et~al.} 2010, \aj,
  140, 1868

\bibitem[{{Xu} {et~al.}(2009){Xu}, {Voronkov}, {Pandian}, {Li}, {Sobolev},
  {Brunthaler}, {Ritter}, \& {Menten}}]{xu09}
{Xu}, Y., {Voronkov}, M.~A., {Pandian}, J.~D., {et~al.} 2009, \aap, 507, 1117

\bibitem[{{Yun} {et~al.}(2009){Yun}, {Elia}, {Palmeirim}, {Gomes}, \&
  {Martins}}]{yun09}
{Yun}, J.~L., {Elia}, D., {Palmeirim}, P.~M., {Gomes}, J.~I., \& {Martins},
  A.~M. 2009, \aap, 500, 833

\end{thebibliography}
\end{document}